\documentclass[10pt,final,journal]{IEEEtran}
\usepackage[pdftex]{graphicx}
\usepackage[cmex10]{amsmath}
\usepackage{amssymb}
\usepackage[colorlinks, bookmarks=false, citecolor=blue, linkcolor=red, urlcolor=blue]{hyperref}
\usepackage{pifont}

\usepackage{bm}

\begin{document}

\title{Disturbance propagation, inertia location and slow modes in large-scale high voltage power grids}
\author{Laurent~Pagnier,~\IEEEmembership{Member,~IEEE} and Philippe~Jacquod,~\IEEEmembership{Member,~IEEE}}
\maketitle

\begin{abstract}
Conventional generators in power grids are steadily substituted with new renewable sources of electric power. 
The latter are connected to the grid via inverters and as such have little, if any rotational inertia. The resulting 
reduction of total inertia raises important issues of 
power grid stability, especially over short-time scales. We have constructed a
model of the synchronous grid of continental Europe with which we numerically 
investigate frequency deviations as well as rates of change of frequency (RoCoF)
following abrupt power losses. The  magnitude of
RoCoF's and frequency deviations  strongly depend on the fault location, and we find the largest effects 
for faults located on the slowest mode -- the Fiedler mode -- 
of the network Laplacian matrix. This mode
essentially vanishes over Belgium, Eastern France, Western Germany, northern Italy and Switzerland. Buses inside these regions 
are only weakly affected by faults occuring outside. Conversely, faults inside these regions have only a local effect
and disturb only weakly outside buses. 
Following this observation, we reduce rotational inertia through three different procedures by either (i)
reducing inertia on the Fiedler mode, (ii) reducing inertia homogeneously and (iii) reducing inertia outside
the Fiedler mode. We find that procedure (iii) has little effect on disturbance propagation, 
while procedure (i) leads to the strongest increase 
of RoCoF and frequency deviations. These results for our model of the European transmission grid are
corroborated by numerical investigations on the ERCOT transmission grid. 
\end{abstract}


\section{Introduction}\label{section:intro}

The short-time voltage angle and frequency dynamics of AC power grids is standardly modeled by the 
swing equations~\cite{mac08}. The latter determine how local disturbances about the 
synchronous operational state propagate through the grid. They emphasize in particular how voltage 
angle and frequency excursions are partially absorbed on very short time scales by the inertia of rotating machines, before 
primary control sets in. With the energy transition, more and more new renewable energy sources (RES) such as 
solar photovoltaic units -- having no inertia -- and wind turbines -- whose inertia is at this time essentially
suppressed by inverters  --  substitute for conventional power generators. 
The resulting overall reduction in rotational inertia 
raises a number of issues related to system dynamics and stability~\cite{ulb14,tie16}. 
It is in particular desirable to determine how much inertia is sufficient and where to optimally locate it
to guarantee short-time grid stability. Determining the optimal placement of inertia is of paramount importance 
at the current stage of the energy transition, as it would help determine where the substitution of  
conventional generators by RES crucially needs to be accompanied by the deployment of 
synchronous condensers or synthetic inertia. 

The impact of lowered levels of inertia on grid stability has been investigated in a number of papers. 
Ulbig et al.  investigated the impact of reduced inertia on 
power system stability for a two-area model~\cite{ulb14}. Extended to three-area systems their analysis led them
to postulate that, at fixed amount of inertia, meshed grids have a greater resilience to disturbances than unmeshed ones~\cite{ulb15}. 
These works further raised the issue of optimal inertia placement in a grid with reduced total amount of inertia. This issue 
is interesting from the point of view of synthetic inertia, obtained by controlling the inverters connecting RES to the 
grid~\cite{bev14,yan17} and which can in principle be deployed where needed. It is moreover crucial to 
anticipate where the substitution of conventional power generators would require significant inertia compensation and where
not. Borsche et al. evaluated 
damping ratios and transient overshoots to optimize the placement of virtual inertia~\cite{bor15}.  Poolla et al. proposed
a different placement optimization based on the minimization of ${\cal H}_2$ norms~\cite{poo17}, while Pirani et al. adopted an
approach based on  ${\cal H}_\infty$ norms~\cite{pir17}. As pointed out by Borsche and D\"orfler~\cite{bor18},
the objective functions to be minimized in these works are not directly related to the standard operational 
criteria of Rate of Change of  Frequency (RoCoF) or frequency deviations in electric power grids. To bridge that gap, 
Ref.~\cite{bor18} constructed an inertia placement optimization algorithm based on these criteria.

In this paper we investigate RoCoF's under abrupt power losses in high voltage
power grids. Our goal is to understand
how the ensuing disturbance propagates through the system, as a function of the power fault location. 
To that end we construct a
model of the synchronous high voltage grid of continental Europe that includes geolocalization, dynamical parameters and 
rated voltage of all buses, as well as electrical parameters of all power lines. 
Our approach is mostly numerical and therefore is not limited by assumptions of constant inertia or damping coefficients 
that are necessary to obtain analytical results. Our model is unique in that it is based on a realistic map of the distribution of 
rotational inertia in the synchronous European grid. 
Disturbance propagation under noisy perturbations have been investigated in 
a number of works on dynamical networks (see e.g. Ref.~\cite{sia16,wol17,tam18}). What makes the present work special is
the spatio-temporal resolution of our investigations on large-scale networks,
which allows us to correlate the impact of the location of the fault with the 
nonhomogeneous distribution of inertia and the spatial support of the slowest modes of the network Laplacian. 

\begin{figure*}[h!]
\center
\includegraphics[width=\textwidth]{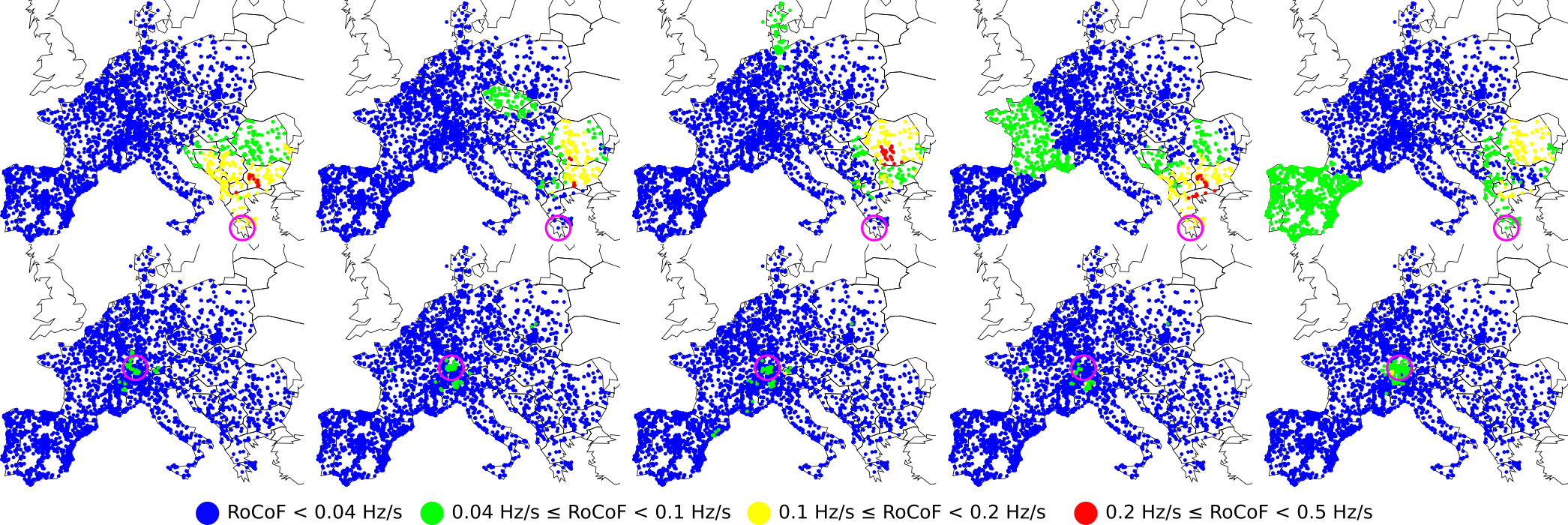}
\caption{Spatio-temporal evolution of local RoCoFs for two different power losses of $\Delta P=900$ MW in a moderate load (typical of a standard summer evening) configuration of the synchronous grid of continental Europe of 2018. 
The top five panels correspond to a fault in Greece and the bottom five to a fault in Switzerland. In both cases, 
the fault location is indicated by a purple circle. Panels correspond to 
snapshots over time intervals 0-0.5[s], 0.5-1[s], 1-1.5[s], 1.5-2[s] and 2-2.5[s] from left to right.}\label{fig:RoCoF_snapshots}
\end{figure*}

We numerically simulate sudden power losses at
different locations on the grid for various loads. Using
the swing equations, we evaluate how the resulting frequency disturbance propagates through the grid by recording 
RoCoF's at all buses.
This is illustrated in Fig.~\ref{fig:RoCoF_snapshots} for two different fault locations of the same magnitude, $\Delta P = 900$ MW.
This relatively moderate fault (on the scale of the European grid) 
generates a significant response, with RoCoF's reaching 0.5 Hz/s, over large areas  for a
power loss in Greece. On the other hand, RoCoF's never exceed 0.1 Hz/s when a fault of the same magnitude 
occurs in Switzerland. After a systematic 
investigation of faults over the whole grid, we relate these differences in behavior to (i) the local inertia density in the 
area near the fault at times $t\lesssim 1-2$ s and (ii) the amplitude of the slowest
modes of the grid Laplacian on the faulted bus for  times $t \gtrsim 1-2$ s. 
Point (i) is expected and well known, however point (ii) is, to the best of our knowledge, a new observation. It has in particular 
the surprising consequence that, when the slowest modes have disconnected structures as is the case of the European grid, 
frequency disturbances propagate between distant areas without affecting areas in between.
Comparing different scenarios for inertia withdrawal, corresponding to substituting new RES
for conventional power plants in different regions, we find that inertia withdrawal from areas with large components of the 
slowest modes of the grid Laplacian results in significantly higher RoCoF's.
This has important consequences for planning and optimal inertia location in future low-inertia power grids. 

This manuscript is organized as follows. In Section~\ref{section:power_sys_dynamics} we give some details of 
our model and approach. Section~\ref{section:local_RoCoF} presents numerical investigations of RoCoF's under
abrupt power losses at different locations on the power grid. It relates the magnitude of the response to such faults to 
the location of the fault, in particular the amplitude on the slowest Laplacian modes on the faulted bus. 
Section~\ref{section:reduction} amplifies on Section~\ref{section:local_RoCoF} by investigating the effect 
of reducing the inertia on different areas of the grid. Section~\ref{section:conclusion} summarizes our conclusions.
Details on the model and further numerical results are presented in the Appendix.

\section{Transmission grid model}\label{section:power_sys_dynamics}

We have imported and combined publicly available data to construct a geolocalized model of the high voltage synchronous grid of 
continental Europe. The geographical location and the electrical parameters of each bus is determined, including voltage level, 
dynamical parameters (inertia and damping coefficients), generator type and 
rated power. Line capacities are extracted from their length. They are
compared with known values for a number of lines and found to be in good agreement. Different load situations are 
investigated using a demographically-based distribution of national loads, together with a dispatch based on a DC 
optimal power flow. Details of these procedures are given in Appendix \ref{section:model}. To confirm our conclusions, 
we alternatively used 
a model of the Texas ERCOT transmission grid~\cite{bir17}, where inertia and damping coefficients are obtained using the 
same procedure as for the European model.

The models are treated within the lossless line approximation~\cite{mac08}, where 
the electrical power $P_i^{\rm e}$ injected or extracted at bus \#$i$ is related to the voltage phase
angles $\{\theta_i\}$ as
\begin{equation}\label{eq:pflow}
P_i^{\rm e}=\sum_{j\in\mathcal{V}}B_{ij} \, V_i V_j \, \sin\big(\theta_i-\theta_j\big)\, .
\end{equation}
Here, $B_{ij}$ gives the imaginary part of the admittance of the power line connecting bus \#$i$ at voltage $V_i$ to  bus \#$j$
at voltage $V_j$ and $\mathcal{V}$ is the set of the $N$ buses in the system. Voltages are assumed constant,
$V_i=V_i^{(0)}$ and are equal to either 220 or 380 kV.
We denote
 $\mathcal{V}_{\rm gen}\subset\mathcal{V}$ the subset of buses corresponding to generator buses. 
Their dynamics is described by the  swing equations~\cite{mac08,ber81}
\begin{equation}
m_i\dot\omega_i+d_i\omega_i=P_i^{(0)}-P_i^{\rm e}\,,\; \text{if } i\in \mathcal{V}_{\rm gen}\,,\label{eq:swing1}
\end{equation}
where $\omega_i=\dot\theta_i$ is the local voltage frequency,
$m_i$ is the inertia and 
$d_i$ the damping coefficients of the generator at bus $i$. The complement subset $\mathcal{V}_{\rm load}
= \mathcal{V} \setminus \mathcal{V}_{\rm gen}$ contains inertialess generator or consumer
buses with frequency dependent loads~\cite{ber81} and a dynamics determined by the swing equations~\cite{mac08,ber81}
\begin{equation} 
d_i\omega_i=P_i^{(0)}-P_i^{\rm e}\,,\; \text{if } i\in \mathcal{V}_{\rm load}\, .\label{eq:swing2}
\end{equation}
In \eqref{eq:swing1} and \eqref{eq:swing2}, $P_i^{(0)}$ gives the power input ($P_i^{(0)}>0$) at generator buses or the power output
($P_i^{(0)}<0$) at consumer buses prior to the fault. We consider that \eqref{eq:swing1} and \eqref{eq:swing2} are 
written in a rotating frame with the rated frequency of $\omega_0 = 2 \pi f$ with $f=50$ or 60 Hz, in which case
$\sum_i P_i^{(0)}=0$. 

In order to investigate transient dynamics following a plant outage, we consider 
abrupt power losses $P_i^{(0)} \rightarrow P_i^{(0)}-\Delta P$ with $\Delta P= 900$ MW on the European grid model and 
500 MW on the ERCOT grid model. In both cases, a single plant is faulted and 
only power plants with $P_i  \ge \Delta P$ can be faulted. The 
values of $\Delta P$ are chosen so that many contingencies with different locations 
homogeneously distributed over the whole grid can be investigated. Larger faults 
with significantly larger RoCoF's will be briefly discussed in the Conclusion section.
Frequency changes are then calculated from \eqref{eq:swing1}
and \eqref{eq:swing2}, with initial conditions given by their stationary solution and 
the faulted bus $\# b$ treated as a  load bus with power injection
$P_b=P_b^{(0)}-\Delta P$, vanishing inertia, $m_b = 0$, and unchanged damping coefficient $d_b$.  
This should be considered as our definition of a fault, where for each faulted generator,  
the same amount of power is always lost, together with the full inertia of the faulted generator. 

\section{Disturbance propagation}\label{section:local_RoCoF}

Our numerical data monitor the voltage angle and frequency excursion following an abrupt power loss. 
Fig.~\ref{fig:RoCoF_snapshots} shows two such events with series of snapshots illustrating the propagation 
of the disturbance over the continental European grid during the first 2.5 seconds after the contingency. 
The two events differ only by the location of the power loss. In the top row the faulted power plant is in Greece, while 
in the bottom row it is in Switzerland (fault locations are indicated by purple circles). 
In both instances, the lost power is $\Delta P= 900$ MW and the grid, including
loads and feed-ins, inertia distribution, damping parameters and electrical parameters of all power lines, is the same.

\begin{figure}[b!]
\center
\includegraphics[width=0.49\textwidth]{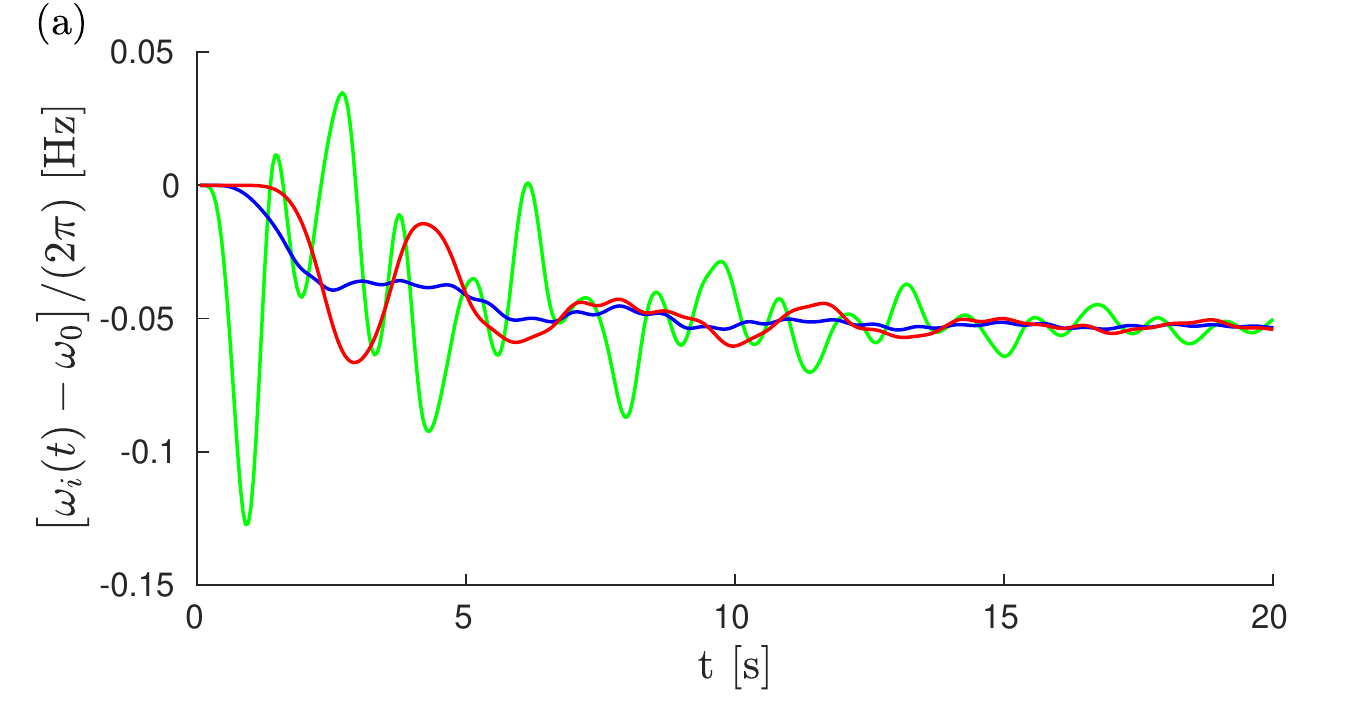}
\includegraphics[width=0.49\textwidth]{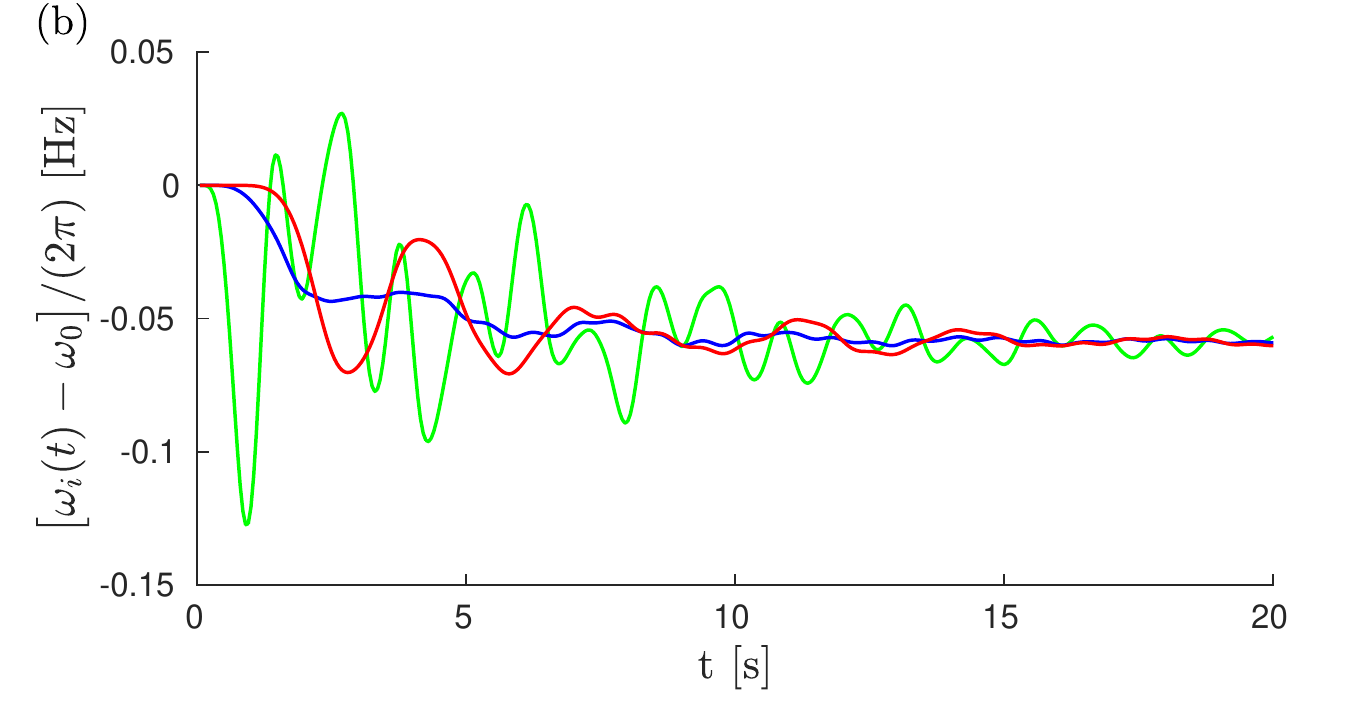}
\caption{Frequency deviations as a function of time for the fault illustrated (a) in the top row of Fig.~\ref{fig:RoCoF_snapshots} 
and (b) in the top row of Fig.~\ref{fig:RoCoF_snapshots2} [with inertia in France reduced by a factor of two compared to 
panel (a)], for 
three buses in the Balkans (green), France (blue) and Spain (red).}\label{fig:freq1}
\end{figure}

The two disturbance propagations shown are dramatically different. For a fault in Greece, 
RoCoF's reach 0.5 Hz/s for times up to 2s. The disturbance furthermore
propagates across almost all of the grid. Quite surprisingly, it seems to jump from Germany to Spain while 
avoiding Eastern France, Belgium and Switzerland in between. 
We have checked that this is not an artifact of the way we plot the average RoCoF, but truly reflects a moderate effect on the
local grid frequency in that area. This is illustrated in
Fig.~\ref{fig:freq1}(a) which shows frequency deviations for three buses
in the Balkans, in Eastern France and in Spain. While the balkanic and spanish buses oscillate rather strongly, the french bus 
displays weak oscillations about a frequency reduction reflecting the loss of power generation $\Delta P$. 
Also remarkable is the RoCoF persistence in eastern Europe at later times, $t>2$ s.

For a fault in Switzerland, on the other hand, RoCoF's never exceed 0.1 Hz/s and the disturbance does not propagate
beyond few hundred kilometers.
We have systematically investigated disturbance propagation for faults located everywhere
on the European grid model and found that major discrepancies between fault located 
in the Portugal-Spain area or the Balkans generate significantly stronger and longer disturbances, propagating 
over much larger distances than faults located in Belgium, Eastern France, Western Germany or Switzerland. 

\begin{figure}[b!]
\center
\includegraphics[width=0.45\textwidth]{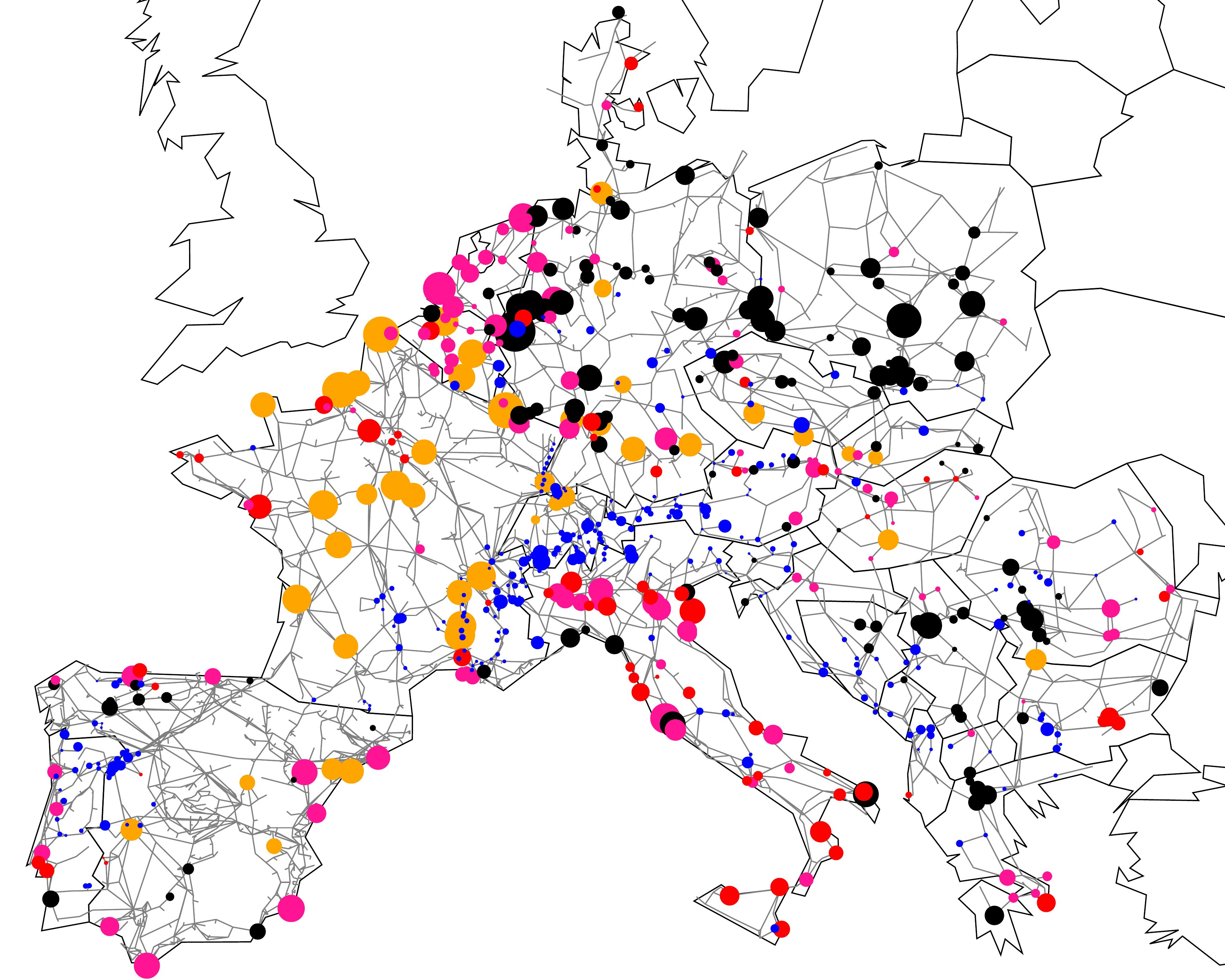}
\caption{Inertia parameters of generators in our model of the synchronous grid of continental Europe. 
The disk size is proportional to $m_i$ and the colors label 
hydro (blue), nuclear (orange), gas (pink), coal (black) and other (red) power plants.}\label{fig:inertia_map}
\end{figure}

\begin{figure*}[h!]
\center
\includegraphics[width=\textwidth]{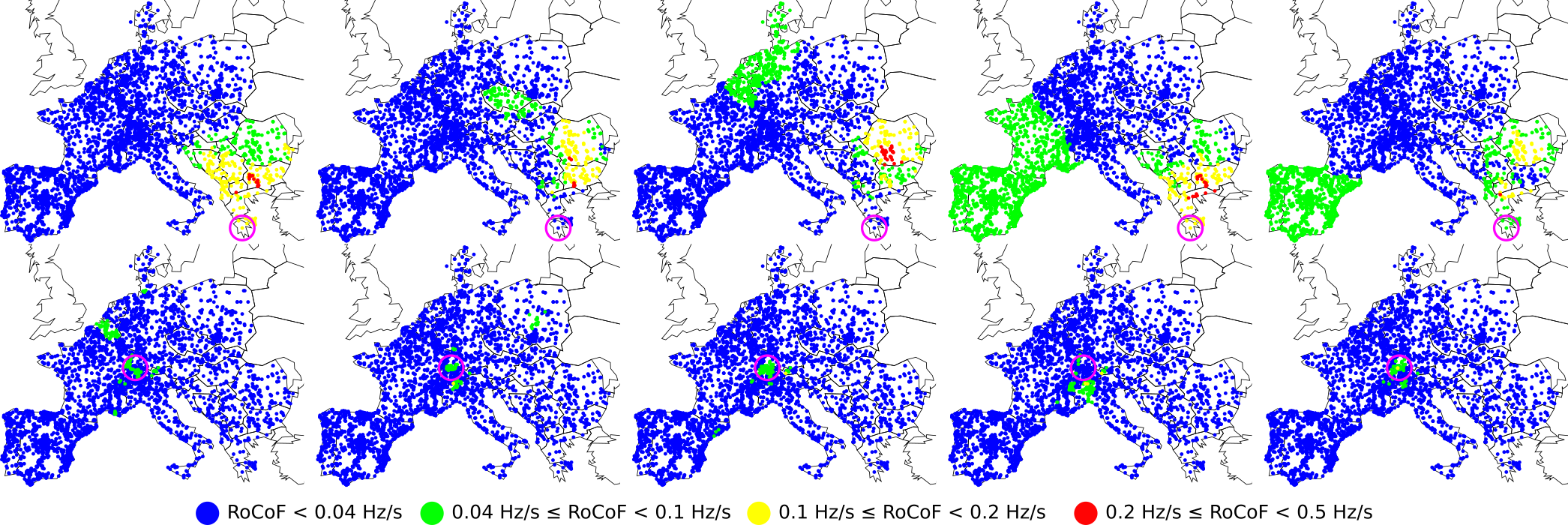}
\caption{Spatio-temporal evolution of local RoCoFs for two different power losses of $\Delta P=900$ MW in a moderate load (typical of a standard summer evening) configuration of the synchronous grid of continental Europe of 2018. 
The top five panels correspond to a fault in Greece and the bottom five to a fault in Switzerland. In both cases, 
the fault location is indicated by a purple circle. Panels correspond to 
snapshots over time intervals 0-0.5[s], 0.5-1[s], 1-1.5[s], 1.5-2[s] and 2-2.5[s] from left to right. The situation is the same as in 
Fig.~\ref{fig:RoCoF_snapshots},
except that all inertia coefficients $m_i$ in France have been divided by two.}\label{fig:RoCoF_snapshots2}
\end{figure*}

This discrepancy in behaviors is partly due to the distribution of inertia in the European grid. As a matter of fact, 
the latter is not homogeneous, as is shown in Fig.~\ref{fig:inertia_map}. Inertia density is smaller in Spain
and Eastern Europe and larger in a strip from Belgium to Northern Italy, including France, Western Germany
and Switzerland. 
Inertia is not only position-dependent, it is also time-dependent as it is directly related to the rotating machines
connected to the grid at any given time~\cite{ulb14,ulb15}. Our results below are obtained both for a typical summer 
evening (with moderate load and thus reduced total inertia) and a typical winter evening (with large load and 
thus larger total inertia). 

\begin{figure*}[h!]
\center
\includegraphics[width=\textwidth]{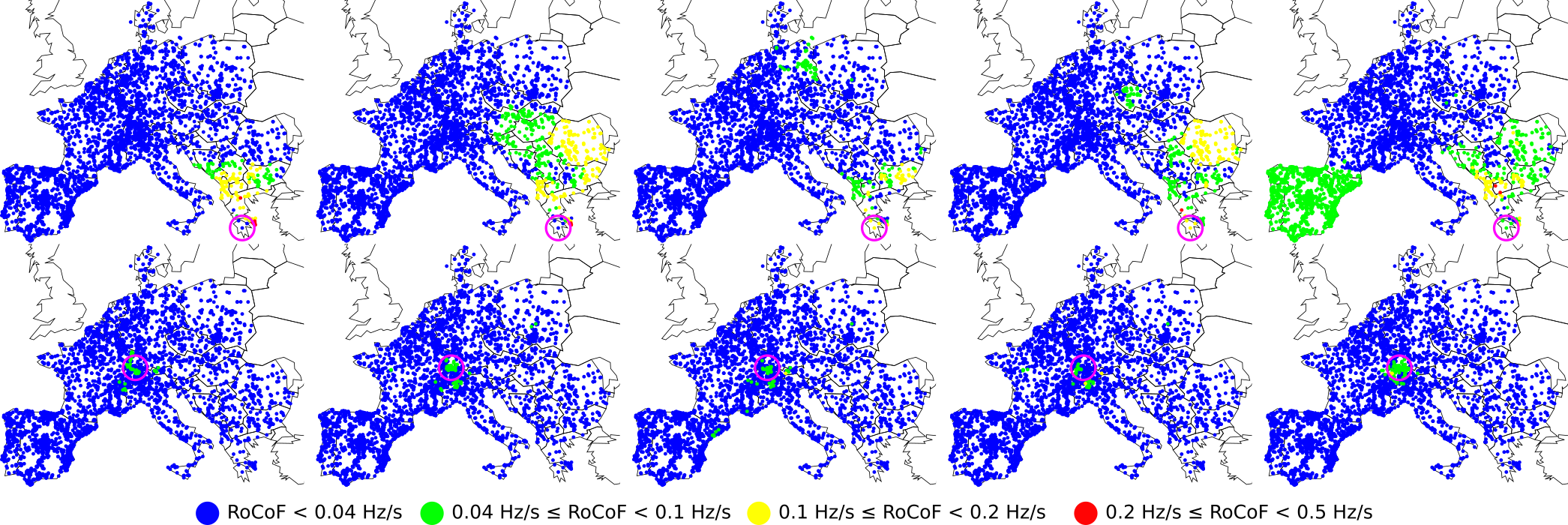}
\caption{Spatio-temporal evolution of local RoCoFs for two different power losses of $\Delta P=900$ MW in a moderate load (typical of a standard summer evening) configuration of the synchronous grid of continental Europe of 2018. 
The top five panels correspond to a fault in Greece and the bottom five to a fault in Switzerland. In both cases, 
the fault location is indicated by a purple circle. Panels correspond to 
snapshots over time intervals 0-0.5[s], 0.5-1[s], 1-1.5[s], 1.5-2[s] and 2-2.5[s] from left to right.
The situation is the same as in
Fig.~\ref{fig:RoCoF_snapshots},
except that all inertia coefficients $m_i$ in the Balkans have been multiplied by two.}\label{fig:RoCoF_snapshots3}
\end{figure*}

To investigate the influence of inertia distribution on frequency disturbance propagation,
we simulated the same faults as in Fig.~\ref{fig:RoCoF_snapshots}, first, artificially reducing inertia by a factor of two 
in France, second artificially increasing inertia by a factor of two in the Balkans. The results are shown 
in Figs.~\ref{fig:RoCoF_snapshots2} and \ref{fig:RoCoF_snapshots3} respectively. First, one sees 
in Fig.~\ref{fig:RoCoF_snapshots2} that reducing the 
inertia in France has only a local effect. Frequency disturbance from a nearby fault in Switzerland propagates over a
larger distance with reduced inertia in France, as expected, 
however there is very little effect on disturbance propagation from a power loss
in Greece. Particularly interesting is that even with a two-fold reduction of inertia in France, there is no increase of the
disturbance affecting Eastern France and only a mild increase of it in Western France from a power loss in Greece. 
Frequency deviations for three buses
in the Balkans, in Eastern France and in Spain are shown in Fig.~\ref{fig:freq1}(b). 
Despite the reduced inertia in France, the french bus still
displays only weak oscillations of frequency about an average frequency decrease characteristic 
of power losses. The overall frequency evolution is surprisingly almost indistinguishable from the case with normal inertia in 
France in Fig.~\ref{fig:freq1}(a). 
Second, increasing the inertia in the Balkans certainly absorbs
part of the  frequency disturbance  from a nearby fault in Greece. This can be seen in Fig.~\ref{fig:RoCoF_snapshots3}.
However, relatively large RoCoF's still persists at  $t>2$ s,
with a magnitude that is reduced only because the initial disturbance has been partially absorbed by the increased
inertia at short times, $t<1$ s.
We therefore conclude that strong discrepancies in frequency disturbance propagation depending on the location 
of an abrupt power loss cannot be understood solely on the basis of inertia distribution. In particular, (i) it is not inertia that
renders France almost immune to frequency disturbance generated by a power loss in Eastern Europe or Spain, 
(ii) it is not only the lack of inertia in the Balkans that allows
the persistance there of relatively large RoCoF's at $t > 2$s.  
Figs.~\ref{fig:RoCoF_snapshots4} and \ref{fig:RoCoF_snapshots5} in the Appendix further show similar behaviors
in disturbance propagation following different faults or under different initial load configurations. 

We can gain some qualitative understanding into these phenomena through spectral graph theory under simplifying assumptions
on model parameters. From
\eqref{eq:swing1} and \eqref{eq:swing2} and for small angle differences between connected buses in the initial 
operational state, weak angle deviations have a dynamics governed by
\begin{eqnarray}\label{eq:linearswing}
{\bm M}\dot{{\bm \omega}} +{\bm D} {\bm \omega} = {\bm P} - {\bm L} \, {\bm \theta} \, ,
\end{eqnarray}
with the diagonal matrices ${\bm M}=\textrm{diag}(\{m_i\})$ ($m_i \ne 0$ for generators only), $\bm D = \textrm{diag}(\{d_i\})$
and the Laplacian matrix ${\bm L}$ of the grid, with elements $({\bm L})_{ij}=-B_{ij} V_i^{(0)} V_j^{(0)}$, $i \ne j$ 
and $({\bm L})_{ii}=\sum_k B_{ik}V_i^{(0)} V_k^{(0)}$. Voltage angles and frequencies as well as power injections have been 
cast into vector form in \eqref{eq:linearswing}, i.e. 
$ {\bm \theta}=( \theta_1, ...  \theta_N)$ and so forth. The Laplacian matrix is real and symmetric, as such 
it has a complete orthogonal set of eigenvectors $\{{\bm u}_1,\ldots,{\bm u}_{N}\}$ 
with eigenvalues $\{\lambda_1,\ldots\lambda_N\}$.
The zero row and column sum property of $\bm L$ implies that $\lambda_1=0$, corresponding to an eigenvector
with constant components, $({\bm u}_1)^\top=(1,\ldots,1)/\sqrt{N}$. If the grid is connected, as in our case, 
all other eigenvalues of $\bm L$ are strictly positive, $\lambda_\alpha>0$ for $\alpha=2,\ldots,N$. This guarantees linear
stability of fixed point solutions to \eqref{eq:linearswing} in the sense that small
angle and frequency deviations are exponentially damped with time. 

Our initial state is a stationary state of \eqref{eq:linearswing}, characterized by $\omega_i=0$ $\forall i$ (since we work in the
rotating frame) and a set of voltage angles $\bm \theta^{(0)}$. 
We then let this initial state evolve under \eqref{eq:linearswing} after a power loss with $P_b \rightarrow P_b -\Delta P$
at bus \# $b$.
Using the method of Ref.~\cite{tyl18a,tyl18b} we can compute 
frequency deviations at bus $\#i$ and time $t$ as a spectral sum over the eigenvectors, $\{ \bm u_\alpha\}$, and eigenvalues,
$\{ \lambda_\alpha \}$ of the Laplacian matrix,
under the assumption of homogeneous inertia and damping coefficients, $m_i=m$ and $d_i=d$. One gets, with $\gamma=d/m$,
\begin{align}\label{eq:domega}
\delta \omega_i(t) &= \frac{\Delta P \, e^{-\gamma t/2}}{m} \sum_{\alpha=1}^N u_{\alpha i}u_{\alpha b}\frac{\sin \Big( \sqrt{\lambda_\alpha/m-\gamma^2/4} \, t\Big)}{  \sqrt{\lambda_\alpha/m-\gamma^2/4}} \, .
\end{align}

\begin{figure}[h!]
\center
\includegraphics[width=0.45\textwidth]{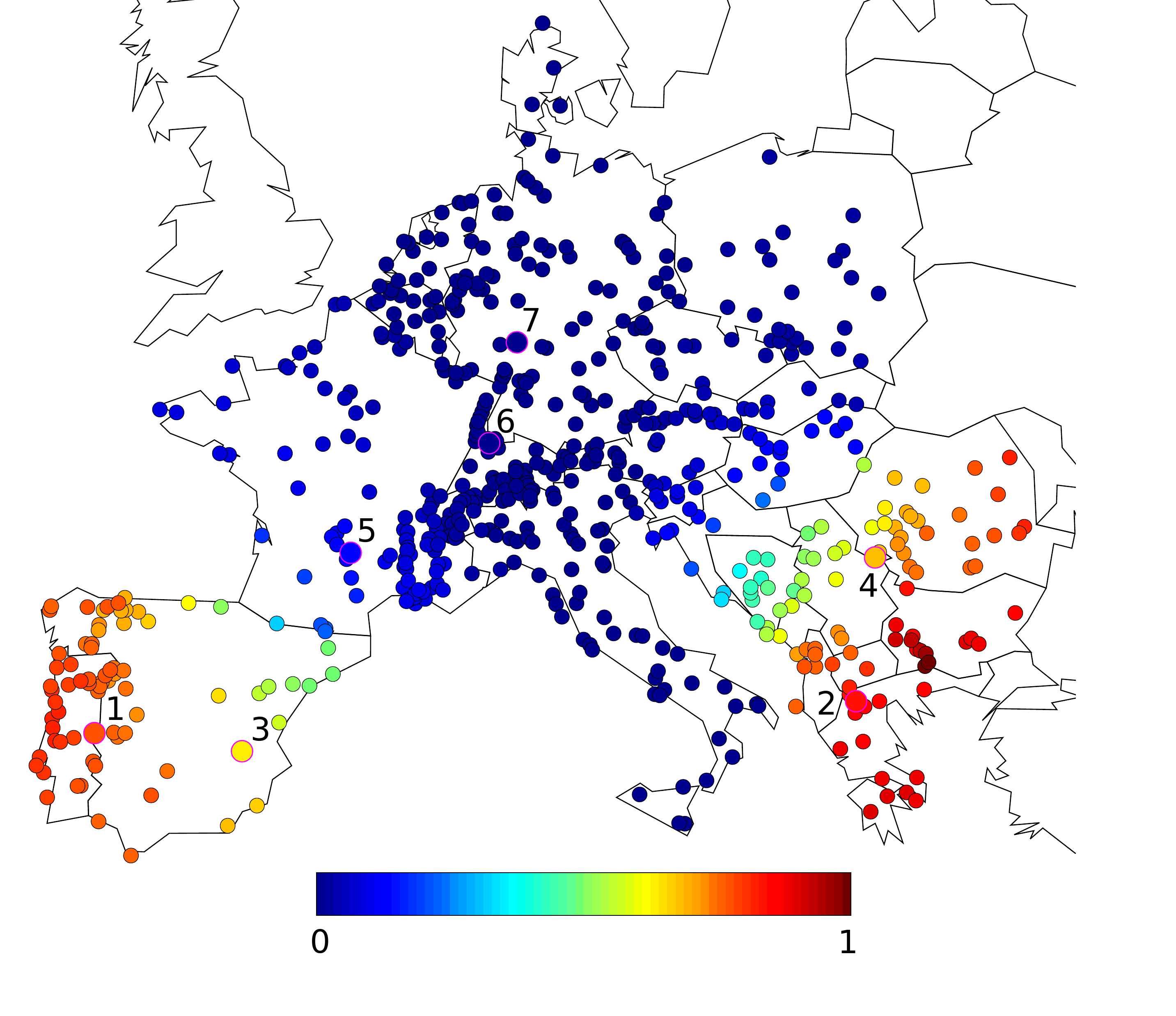}\\
\includegraphics[width=0.35\textwidth]{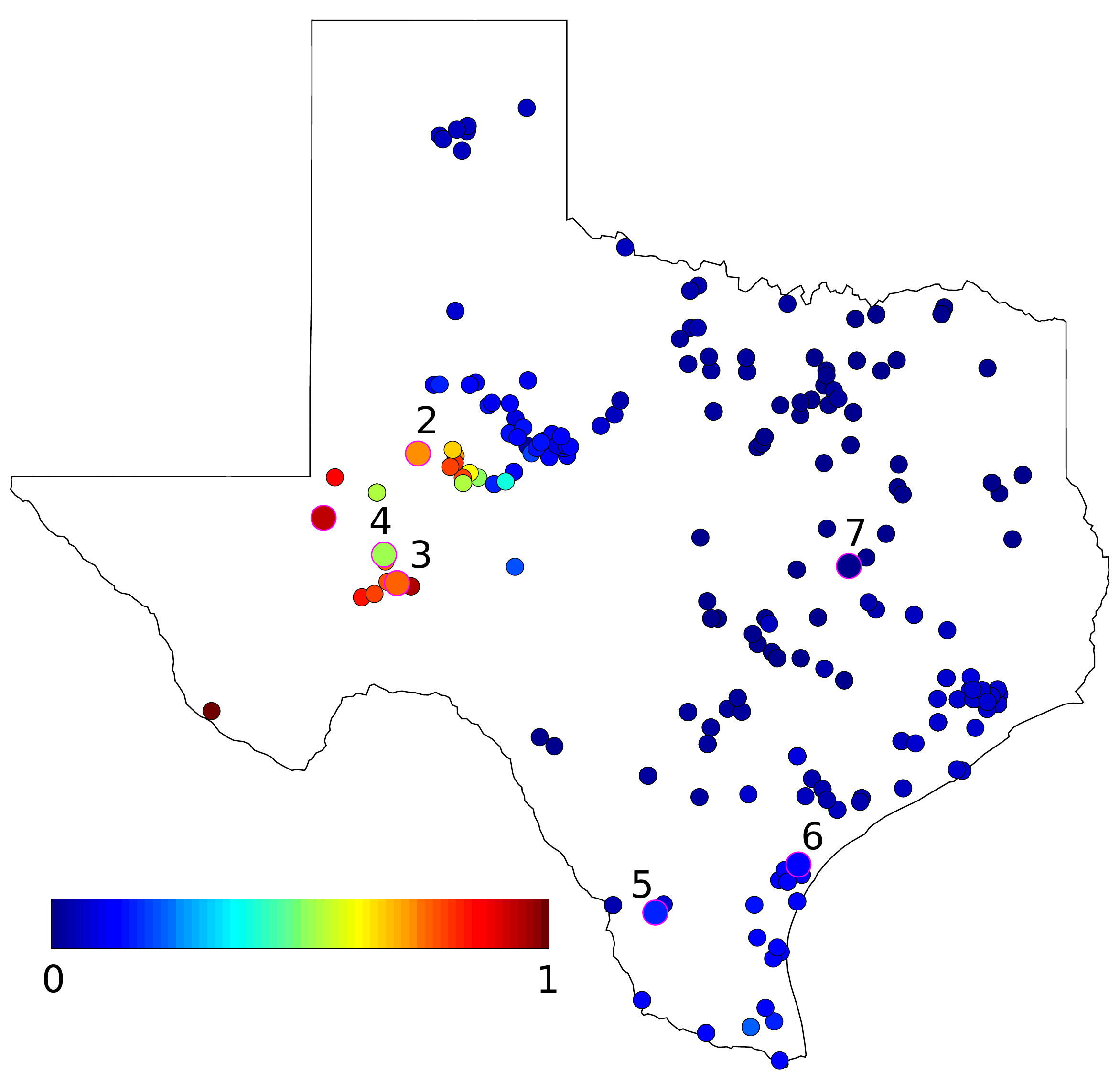}\\
\caption{Color plot of the normalized squared components $u_{2k}^2$ of the Fiedler mode 
on generator buses in the European and ERCOT grids.
Labeled buses correspond to  labeled symbols in Figs.~\ref{fig:m_vs_u} and \ref{fig:uniform_reduction}.}\label{fig:fiedler}
\end{figure}

In real power grids frequencies are monitored at discrete time intervals $t \rightarrow k \Delta t$,
with $\Delta t$ ranging between 40 ms and 2 s~\cite{jen00}.
RoCof's are then evaluated as the frequency slope between two such measurements. 
In our numerics, we use $\Delta t = 0.5$ s.
The RoCoF at bus \#$i$ reads
\begin{equation}\label{eq:rocof2}
r_{i}(t)=\frac{\delta \omega_i\big(t+\Delta t\big)-\delta \omega_i\big(t\big)}{2\pi\Delta t}\,.
\end{equation}
Together with \eqref{eq:domega}, this gives
\begin{eqnarray}\label{eq:rocof}
r_{i}(t)&=&\frac{\Delta P \, e^{-\gamma t/2}}{2 \pi m} \sum_{\alpha=1}^N \frac{u_{\alpha i}u_{\alpha b}}{  \sqrt{\lambda_\alpha/m-\gamma^2/4} \, \Delta t}
 \nonumber \\
& \times &  \left[ e^{-\gamma \Delta t/2} \, \sin \Big( \sqrt{\lambda_\alpha/m-\gamma^2/4} \, (t+\Delta t)\Big) \right. \nonumber \\ 
&& \left. - \sin \Big( \sqrt{\lambda_\alpha/m-\gamma^2/4} \, t\Big) \right] \, .
\end{eqnarray}
The term $\alpha=1$ gives a position-independent contribution to the RoCoF, $r_i^{(1)}(t) =
\Delta P \, e^{-\gamma t} (1-e^{-\gamma \Delta t}) / 2 \pi m N \gamma \Delta t$. It is maximal 
and inversely proportional to the inertia coefficient $m$ at short times,
$r_i^{(1)}(t \rightarrow 0) \rightarrow \Delta P / 2 \pi m N$. 
All other terms $\alpha >1$ have oscillations
with both amplitude and period depending on $\sqrt{\lambda_\alpha/m-\gamma^2/4}$. High-lying eigenmodes with large
$\alpha$ and large eigenvalues $\lambda_\alpha$ therefore contribute much less than low-lying eigenmodes, both
because their oscillation amplitude is reduced and because they oscillate faster, which leads to faster cancellation of terms. 
With our choice of $\Delta t=0.5$ s we find 
$\sqrt{\lambda_\alpha/m-\gamma^2/4}\Delta t\in[0.54,416]$ for $\alpha>1$ in our model. The second lowest value is
$\sqrt{\lambda_3/m-\gamma^2/4}\Delta t=0.89$,
almost twice larger than the first one.
One expects that only few eigenmodes of the network Laplacian, corresponding to its lowest nonvanishing
eigenvalues, effectively matter in the spectral sum in \eqref{eq:rocof}.
Higher-lying modes have only short-lived contributions.

These results show that, for homogeneous inertia and damping, the short-time RoCoF response $r_{i}(t)$  
is inversely proportional to 
the inertia. The behavior at longer times is determined by the magnitude of the few slowest 
eigenmodes of the network Laplacian on both the 
perturbation bus [through $u_{\alpha b}$ in \eqref{eq:rocof}] and the bus where the RoCoF is measured [through $u_{\alpha i}$]. 
Despite its neglect of inhomogeneities in inertia and damping, this simple calculation suggests 
that the so far unexplained behaviors observed in our numerical
results in Figs.~\ref{fig:RoCoF_snapshots}, \ref{fig:RoCoF_snapshots2} and \ref{fig:RoCoF_snapshots3} are related to 
slow modes of the network Laplacian. This hypothesis gains further support when looking at the 
structure of the slowest, $\alpha=2$ mode, so-called Fiedler mode
on the generator buses of the European network shown in
Fig.~\ref{fig:fiedler}. The large squared amplitude $u_{2i}^2$ of the 
Fiedler mode on buses $i$ in Spain and the Balkans is consistent with a disturbance propagating from
one to the other of these regions with only minor disturbance on intermediate regions (such as Eastern France in 
Figs.~\ref{fig:RoCoF_snapshots}, \ref{fig:RoCoF_snapshots2} and \ref{fig:RoCoF_snapshots3}). We have found, but do not show, 
that the next, $\alpha = 3$ mode has essentially the same profile of $u_{3i}^2$ as the Fiedler mode. Furthermore, the next 
modes $\alpha=4, \ldots 6$ largely avoid Belgium, France, Western Germany, Northern Italy and Switzerland.
Higher modes have $\sqrt{\lambda_{\alpha > 6}/m-\gamma^2/4}\Delta t > 4 \sqrt{\lambda_2/m-\gamma^2/4}\Delta t$,
accordingly, their contribution
to RoCoF's, \eqref{eq:rocof}, are at least four times smaller and oscillate four times faster than the Fiedler mode.
We therefore neglect them in our qualitative discussions to come. 

To assess the disturbance magnitude of a power loss at bus $\# b$ over frequencies in the whole grid, one needs
to gather information on RoCoF's at different times and locations. We therefore
introduce the  performance measure
\begin{equation}
\mathcal{M}_b=\sum_{k=1}^{N{\rm_{sim}}}\sum_{i}\big|r_{i}(k \Delta t)\big|\,,\label{eq:Mb}
\end{equation}
where $N{\rm_{sim}}=10$ is the number of time intervals $\Delta t=0.5$s considered in our numerics.
Fig.~\ref{fig:freq1} shows that the total time $N{\rm_{sim}} \Delta t=t_{\rm sim}=5$ s considered in our 
numerical calculation of $\mathcal{M}_b$
is set to include major initial oscillations while neglecting oscillations at longer times of little concern for 
power grids. 

\begin{figure}[h!]
\center
\includegraphics[width=0.47\textwidth]{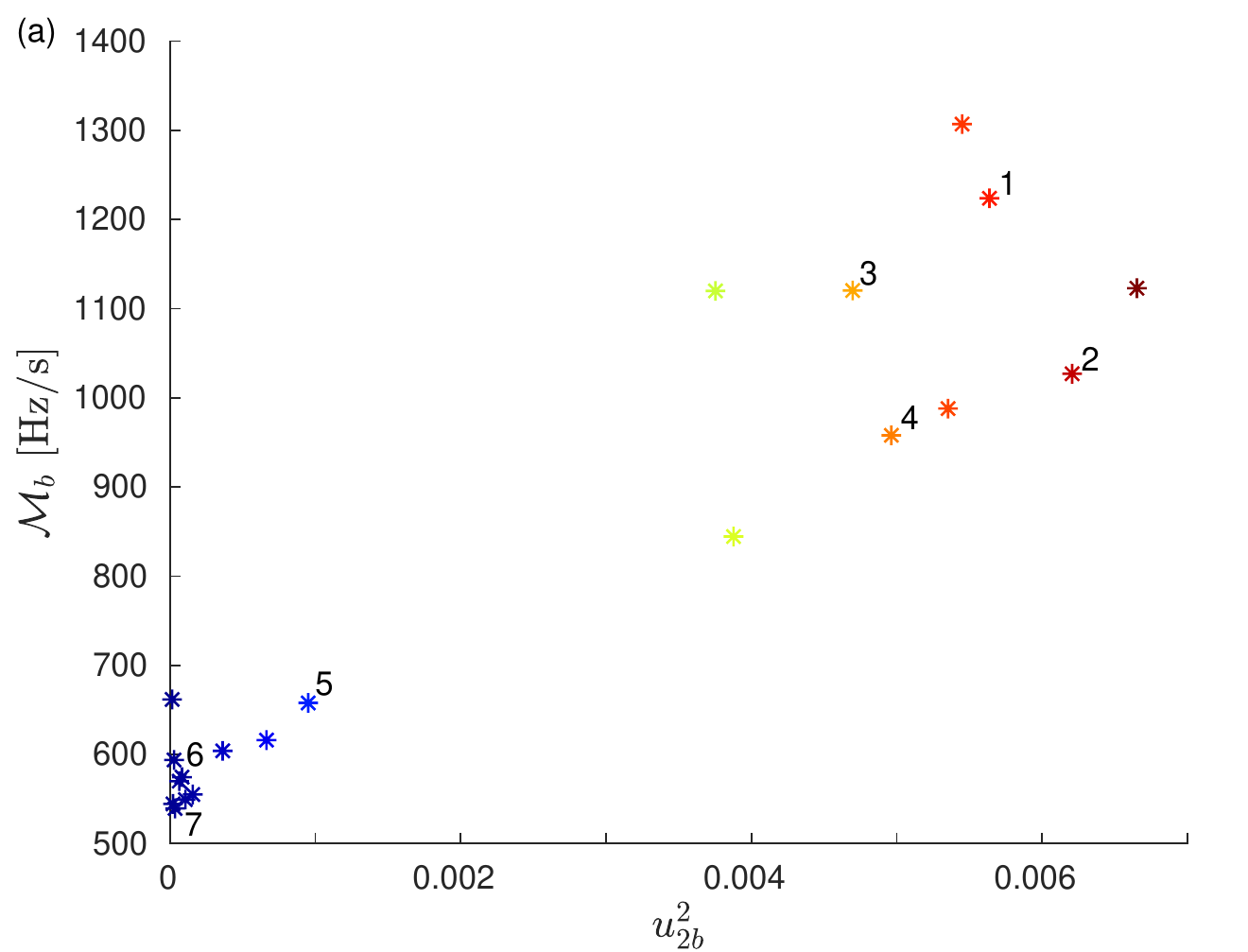}\\
\includegraphics[width=0.47\textwidth]{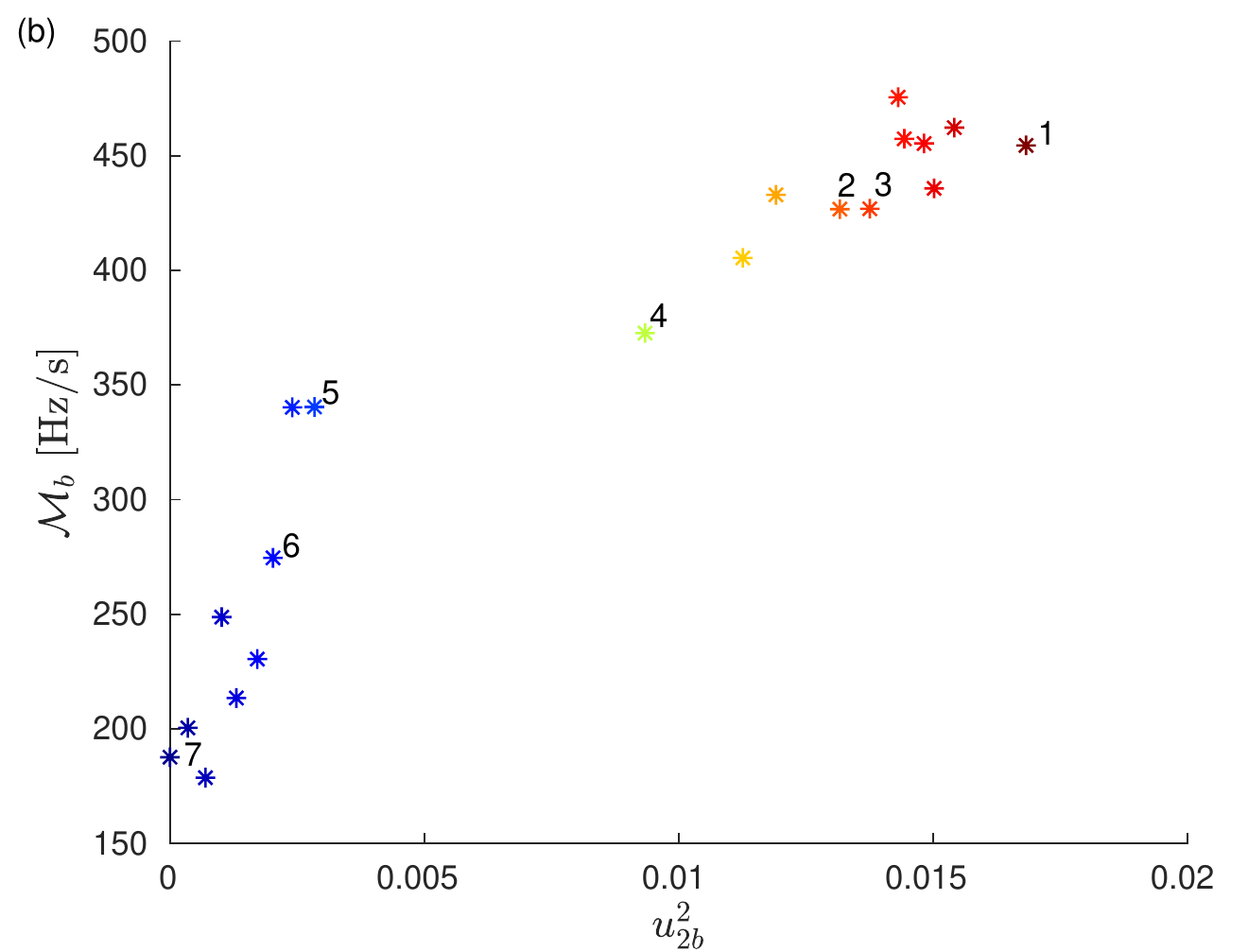}\\
\caption{Global RoCoF disturbance magnitude $\mathcal{M}_b$ as a function of squared Fiedler components $u_{2b}^2$ 
for power losses on 20 different bus \#$b$ for (a) the European and (b) the ERCOT grid.
Labeled symbols correspond to  locations indicated in Fig.~\ref{fig:fiedler}.
The plots look similar when $\mathcal{M}_b$ is plotted against the squared  component $u_{3b}^2$ of the second
slowest mode of the Laplacian.}\label{fig:m_vs_u}
\end{figure}

The above result \eqref{eq:rocof}  suggests that RoCoF's are larger following power losses on 
buses with large components of the eigenvectors with smallest eigenvalues of the Laplacian matrix. To check whether this result
also holds in realistic power grids with nonhomogeneous distribution of inertia, we numerically calculate $\mathcal{M}_b$
for 20 abrupt power losses homogeneously  distributed on the European and ERCOT grids. 
Fig.~\ref{fig:m_vs_u} shows that the disturbance magnitude $\mathcal{M}_b$ grows with 
the squared Fiedler component $u_{2b}^2$ of the location $b$ of the power loss. Everything else being kept constant, 
the disturbance magnitude is more than twice larger
in the European grid and almost three times larger in the ERCOT grid for power losses on 
buses with largest $u_{2b}^2$, than for losses on buses with low $u_{2b}^2$.
As for Fig.~\ref{fig:fiedler}, we have found  that the same trend persists when plotting $\mathcal{M}_b$ against 
the squared component $u_{3b}^2$ of the second slowest mode of the Laplacian. 
The magnitude of the disturbance following an abrupt power loss is therefore determined by
its location, in particular on the amplitude $u_{\alpha b}^2$ on the faulted bus $b$ 
of the Fiedler mode ($\alpha=2$) and of the next 
slowest mode ($\alpha=3$) of the network Laplacian. In what follows, we call, by some abuse of language,
"Fiedler areas" ("non-Fiedler areas") the set of buses $\{ i \}$ where $u_{2i}^2$ and $u_{3i}^2$ are large (small).

\section{Disturbance magnitude vs. inertia}\label{section:reduction}

So far we have 
established that disturbances have strongly location-dependent magnitudes, and in particular that stronger 
disturbances originate from buses with large amplitude of the two slowest eigenmodes of the 
network Laplacian. We next investigate how rotational inertia influences this finding. 
To that end we modify inertia on the network following three different procedures where the inertia of a
generator on bus $\# i$ is increased/decreased according to one of the following probability distributions
\begin{align}
p_{i}^{\rm U}&\propto 1\,,\label{eq:prob_dist1}\\
p_{i}^{\rm F}&\propto u_{2i}^{2}\,,\label{eq:prob_dist2}\\
p_{i}^{\rm nF}&\propto 1/u_{2i}^{2}\,.\label{eq:prob_dist3}
\end{align}
The first procedure reduces/adds inertia uniformly (indicated by the superscript $^{\rm U}$), 
the second one reduces/adds inertia preferentially on buses with large amplitude
of the Fiedler mode (hence the superscript $^{\rm F}$) and the third one reduces/adds inertia
preferentially on buses with small amplitude of the
Fiedler mode (with $^{\rm nF}$ indicating the "non-Fiedler" area). 

\begin{figure}[h!]
\center
\includegraphics[width=0.46\textwidth]{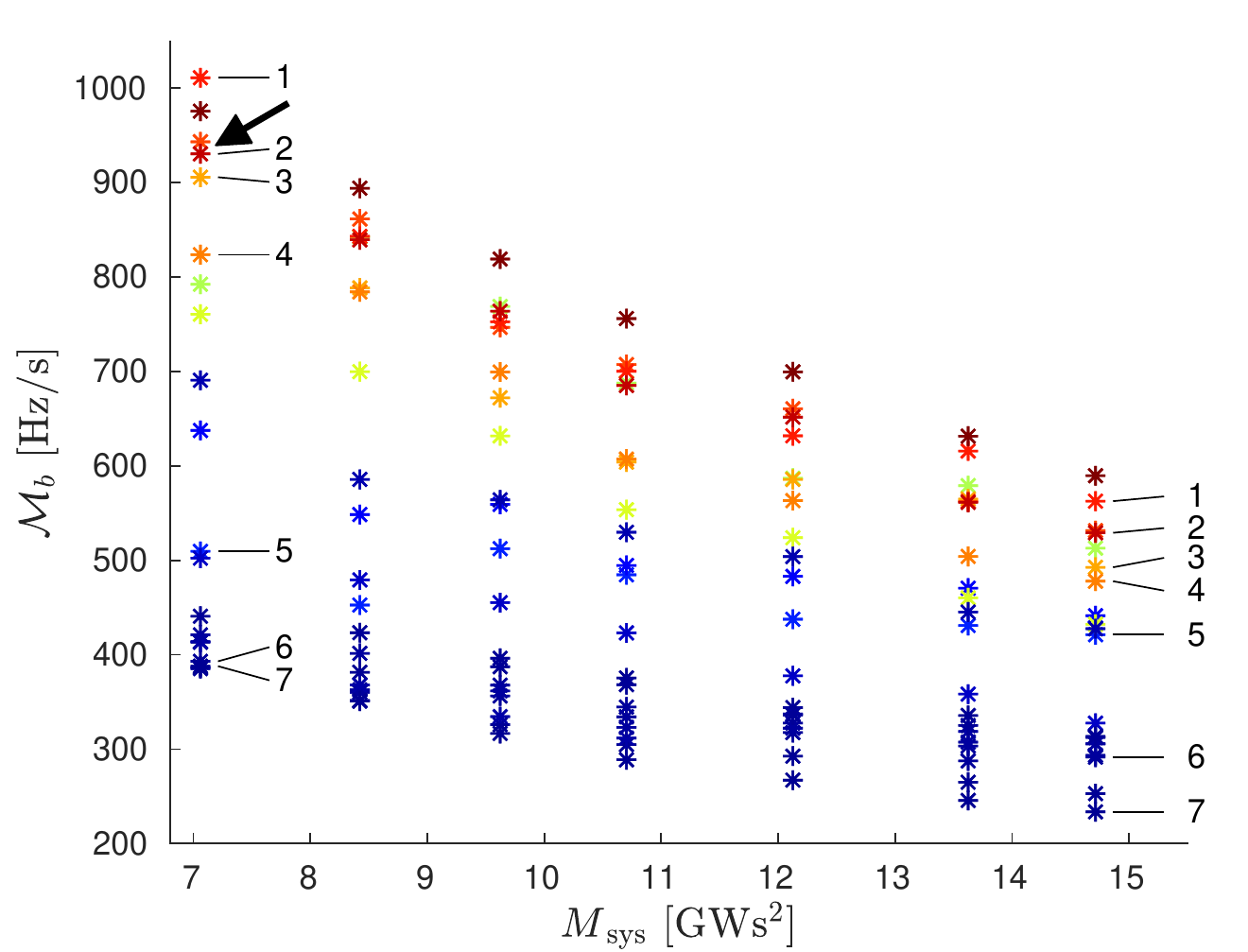}
\caption{Global RoCoF disturbance magnitude $\mathcal{M}_b$ vs. the artificially varied 
system inertia $M_{\rm sys}$ in the European grid. Each point corresponds to the loss of a single power station, 
with colors related to the squared component $u_{2b}^2$ of the Fiedler mode on the power loss bus. Color
code and label symbols are the same as in Fig.~\ref{fig:fiedler}. 
The arrow indicates the data point corresponding to the top left data point, also indicated by an 
arrow in Fig.~\ref{fig:fiedler_path}.
}\label{fig:uniform_reduction}
\end{figure}

Fig.~\ref{fig:uniform_reduction} shows the evolution of $\mathcal{M}_b$ 
as a function of total inertia, $M_{\rm sys}=\sum_{i}m_i,$ for power losses of $\Delta P=900$ MW on the same 
20 power plants as in Fig.~\ref{fig:m_vs_u}. The data corresponding to today's synchronous grid of continental
Europe are the rightmost, with the largest amount of inertia. The inertia is then reduced following the first procedure
where generator buses become randomly  inertialess according to the homogeneous probability distribution \eqref{eq:prob_dist1}.
One sees that $\mathcal{M}_b$ 
follows the ranking defined by the squared Fiedler components, almost regardless of the amount of inertia in the 
system, and 
faults in the Fiedler areas are generically more critical than those in the non-Fiedler areas. 

\begin{figure}[h!]
\includegraphics[width=0.46\textwidth]{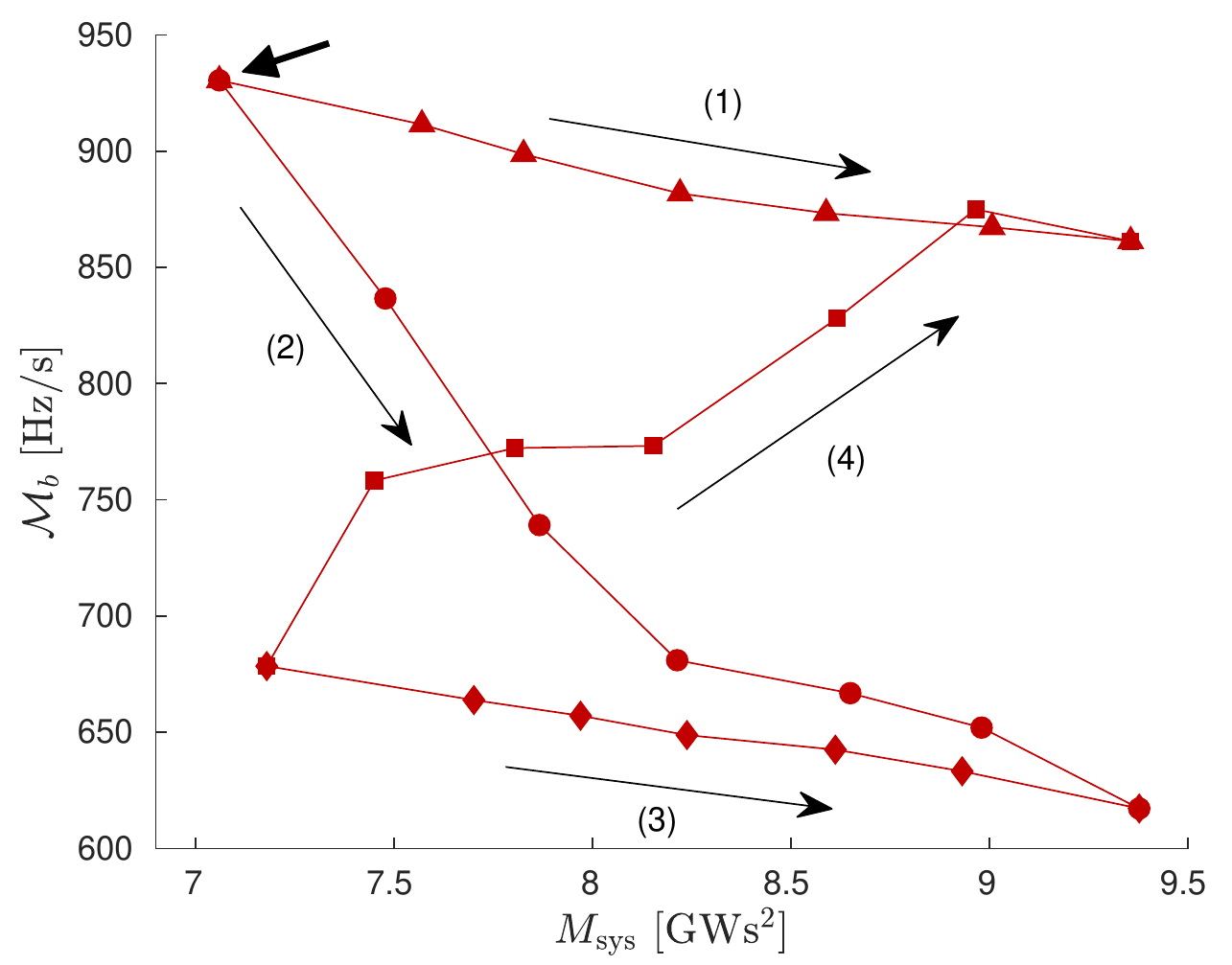}
\caption{Global RoCoF disturbance magnitude $\mathcal{M}_b$ vs. artifically modified total system inertia. 
Along paths (1) and (3), inertia is added according to procedure \eqref{eq:prob_dist3}, i.e. mostly on the non-Fiedler area. 
Path (2) follows procedure \eqref{eq:prob_dist2} by adding inertia almost exclusively on the Fiedler area.
Path (4) follows a selected procedure where inertia is removed from the Fiedler area and added on the non-Fiedler area.
The top left data point corresponds to the data point indicated by an arrow in Fig.~\ref{fig:uniform_reduction}.}\label{fig:fiedler_path}
\end{figure}

The situation can be dramatically different when following other procedures to add/remove inertia selectively on certain areas. 
In Fig.~\ref{fig:fiedler_path}, $\mathcal{M}_b$ is shown, always for the same power
loss. The top left data point (indicated by an arrow) corresponds to the data labeled 2 in the top left of 
Fig.~\ref{fig:uniform_reduction} (also indicated by an arrow). Paths
(1) and (3) correspond to adding inertia according to procedure \eqref{eq:prob_dist3}, i.e. mostly outside the Fiedler area.
This procedure reduces $\mathcal{M}_b$ by less than 10 \% upon increasing the total inertia $M_{\rm sys}$ by 30 \%.
Path (2) follows procedure \eqref{eq:prob_dist2} by adding inertia almost exclusively on the Fiedler area. It is much more
efficient and leads to a reduction of $\mathcal{M}_b$ by more than 30 \% with the same total increase of $M_{\rm sys}$ by 30 \%.
Finally, path (4) illustrates a procedure where inertia is removed from Fiedler areas and added to non-Fiedler areas. 
In that case, the RoCoF disturbance magnitude increases, even with a global increase of inertia. Taken in reverse direction, 
path (4) in Fig.~\ref{fig:fiedler_path} shows that, quite unexpectedly,  
grid resilience against faults such as power losses can be enhanced while simultaneously reducing the total amount of inertia.

\begin{figure}[h!]
\includegraphics[width=0.46\textwidth]{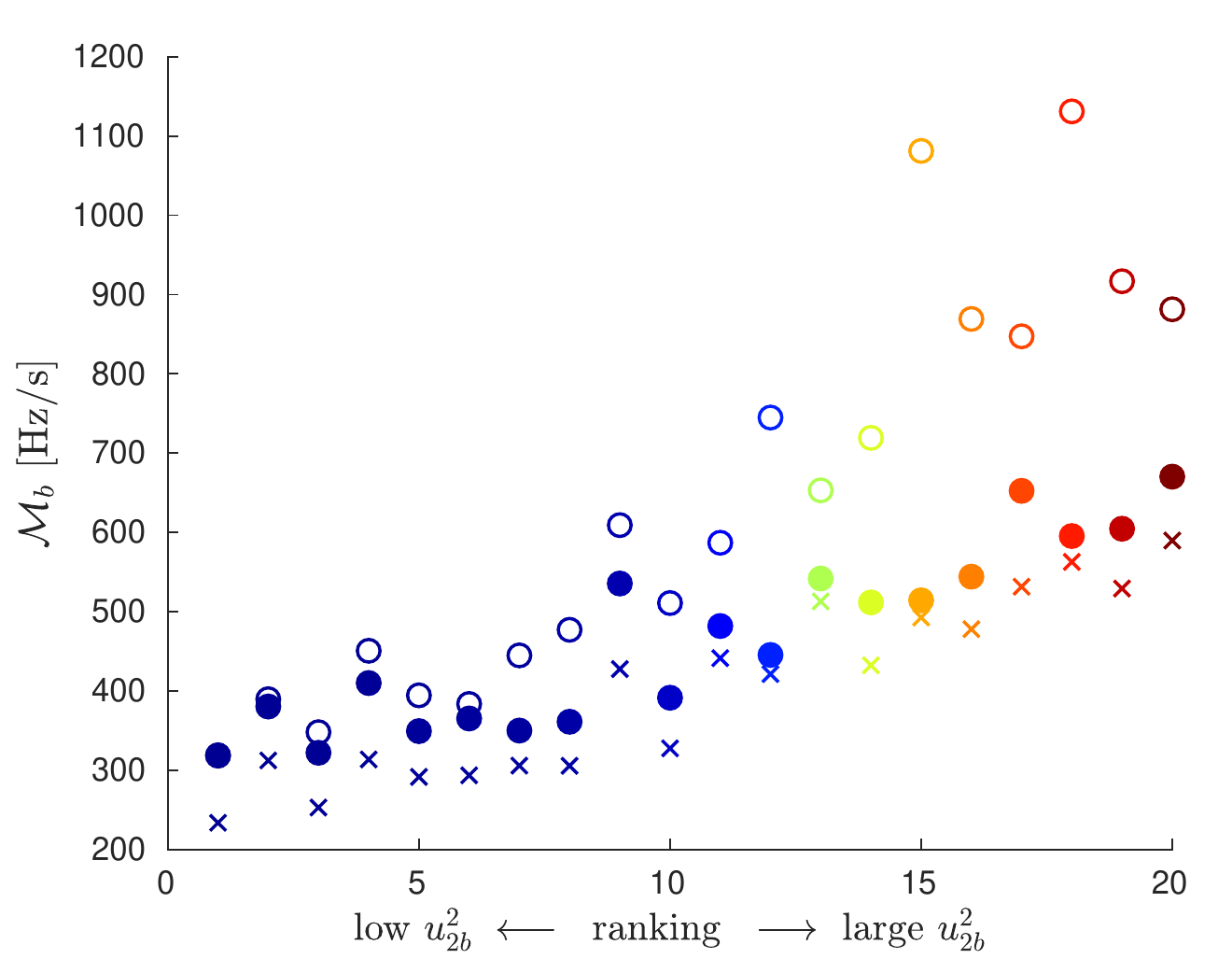}
\caption{Global RoCoF disturbance magnitude $\mathcal{M}_b$ horizontally ranked in increasing order of 
the squared Fiedler mode amplitude $u_{2b}^2$ on the faulted bus \# $b$.
Crosses are for a system with inertia $M_{\rm sys}^0=14.7$ GWs$^2$
(corresponding to today's European grid, Fig.~\ref{fig:RoCoF_snapshots})
and circles for 
reduced inertia $M_{\rm sys}=0.6M_{\rm sys}^0$, with system inertia mainly reduced outside Fiedler areas (solid circles) 
or mainly reduced inside Fiedler areas (empty circles).}\label{fig:dist_msys_fixed}
\end{figure}

We finally show in Fig.~\ref{fig:dist_msys_fixed} how global RoCoF disturbance magnitudes depend on the location of each of the 
20 power losses considered in Fig.~\ref{fig:uniform_reduction}, once inertia is reduced starting from 
 the full inertia situation of Fig.~\ref{fig:uniform_reduction} with $M_{\rm sys}^0=14.7$ GWs$^2$. The three data sets 
 correspond to unchanged inertia $M_{\rm sys}^0$ (crosses), inertia $M_{\rm sys} = 0.6 M_{\rm sys}^0$
 reduced mostly in the Fiedler area, following the probability
 distribution \eqref{eq:prob_dist2} (empty circles) or outside the Fiedler area, according to \eqref{eq:prob_dist3} (full circles).  
 Fig.~\ref{fig:dist_msys_fixed} clearly shows that (i) regardless of the position of the fault, inertia reduction on the Fiedler area
 systematically leads to an enhanced sensitivity to power loss, compared to inertia reduction outside the Fiedler area and (ii)
the sensitivity increase is larger for faults on the Fiedler area. 
 
 \begin{figure*}[h!]
\center
\includegraphics[width=\textwidth]{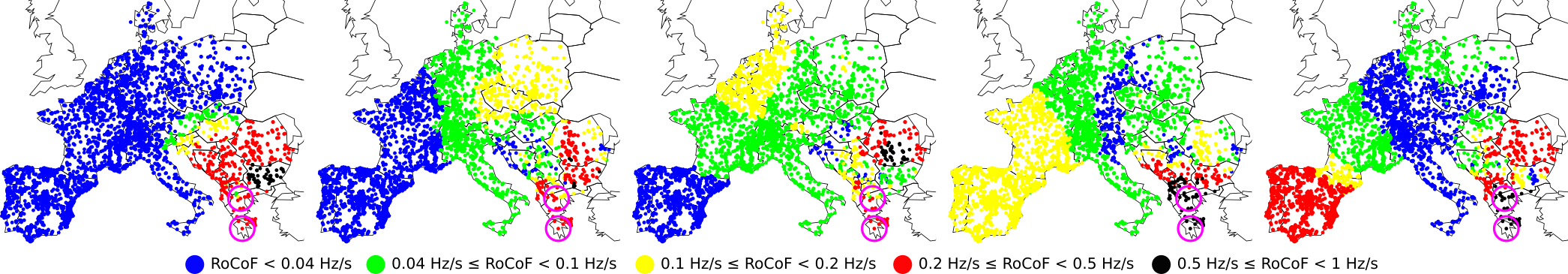}
\caption{Spatio-temporal evolution of local RoCoFs for two simultaneous 
abrupt power losses, each of $\Delta P=1500$ MW in a moderate load (and thus
low inertia, typical of a standard summer evening) configuration of the synchronous grid of continental Europe of 2018. 
The faults location is indicated by purple circles. Panels correspond to 
snapshots over time intervals 0-0.5[s], 0.5-1[s], 1.5-2[s] and 2.5-3[s] from left to right.}\label{fig:RoCoF_snapshots6}
\end{figure*}

\section{Conclusion}\label{section:conclusion}

We have presented numerical investigations on disturbance propagation following a generator fault in 
the synchronous transmission grid of Continental Europe. The first step was to build up a numerical model, including
all necessary parameters to perform dynamical calculations. To the best of our knowledge, no model of this kind
is publicly available. 

In real power grids, protection devices disconnect generators if the RoCoF or frequency deviations
exceed predetermined thresholds. Therefore we based our performance measure on RoCoF and
investigated how the latter evolves in space and time following 
an abrupt power loss, depending on the location of the latter. We have found that disturbances are stronger,
they propagate further and persist longer for faults located on areas supporting significant amplitude of the 
slowest modes of the network Laplacian. In the case of the European grid we found that the two lowest (but nonzero) modes
are particularly important in that respect. They have similar geographical support which we called the "Fiedler area",
because the lowest nonzero mode of a network Laplacian is often called the "Fiedler mode".
Amplifying on those results we found that inertia reduction on the Fiedler area leads to an amplified RoCoF response, while
reducing the inertia on non-Fiedler area has a much weaker effect, with only a moderate increase of RoCoF's.

\begin{figure}[b!]
\center
\includegraphics[width=0.49\textwidth]{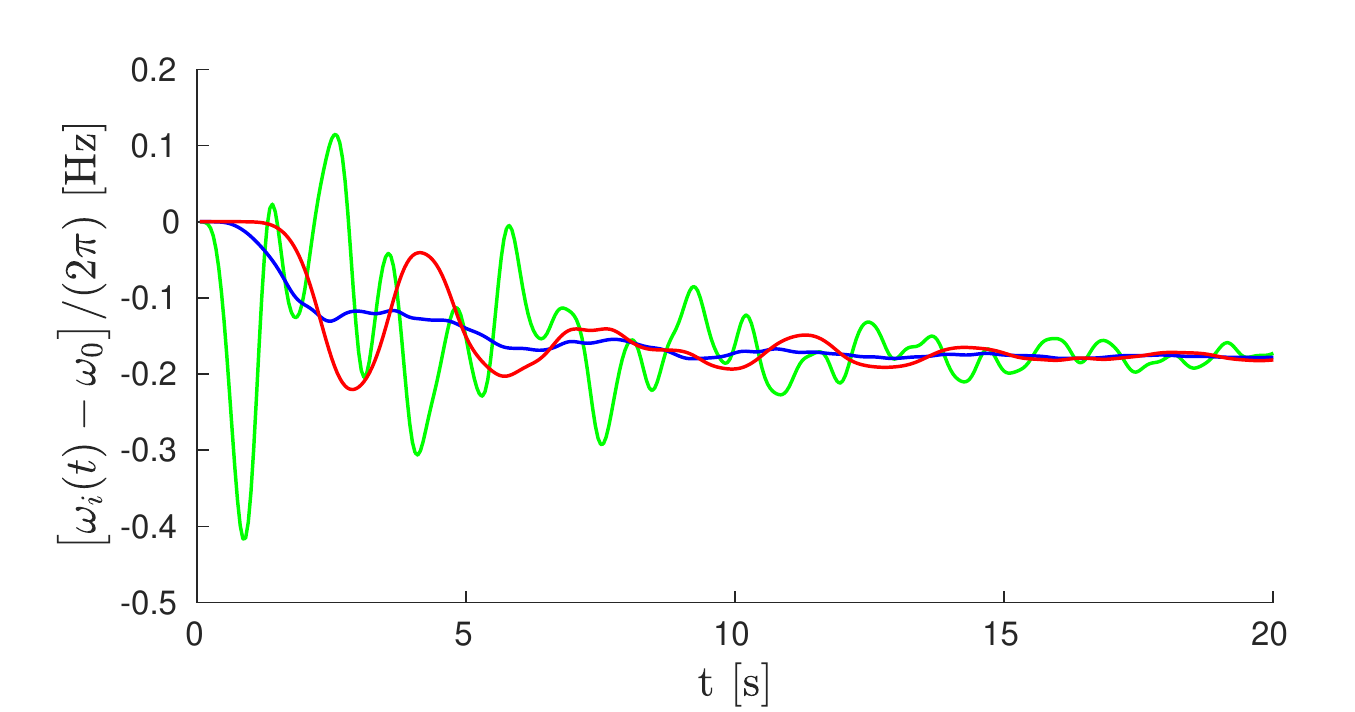}
\caption{Frequency deviations as a function of time for the double fault of $\Delta P=3000$ MW illustrated in Fig.~\ref{fig:RoCoF_snapshots6},
for three buses in the Balkans (green), France (blue) and Spain (red).}\label{fig:freq2}
\end{figure}

The faults considered above correspond to abrupt power losses of $\Delta P=900$ MW. They lead to maximum
RoCoF's magnitudes of 0.5 Hz/s when the fault is located on a Fiedler area under moderate network load conditions.
When the fault is located on a non-Fiedler area, RoCoF's never exceed 0.1 Hz/s. These values are significantly larger when 
considering a normative contingency of $\Delta P=3000$ MW~\cite{entsoe2016frequency}. 
This is usually taken as the tripping of two of the largest generators, connected to the same bus~\cite{entsoe2016frequency},
however since no such generators exist in the Balkans, we show instead
in Fig.~\ref{fig:RoCoF_snapshots6} a similar event resulting from the tripping of two nearby 1500 MW power plants,
for the same load conditions as in Fig.~\ref{fig:RoCoF_snapshots}.
In the ensuing disturbance propagation,
RoCoF's reach values close to 1Hz/s over large areas of south-east Europe for times at least up to 2.5s. Yet, 
even with a fault of this magnitude, the RoCoF's are much weaker in France and other non-Fiedler areas than in the Balkans and 
the spanish peninsula -- where the two slowest modes of the network Laplacian reside. 
Frequency deviations are further shown in Fig.~\ref{fig:freq2} which shows the same qualitative, if not quantitative
behavior as in Fig.~\ref{fig:freq1}, but amplified by the more than three times larger fault magnitude, 
$\Delta P=900$ MW $\rightarrow 3000$ MW. 

Our findings emphasize an important aspect of optimal inertia location. Because long-range 
RoCoF disturbance propagation is controlled by the slowest modes of the network Laplacian,
reducing inertia where these nodes reside, a
areas we called the "Fiedler areas", leads to a significantly more sensitive grid than reducing inertia
outside these areas. Conversely, substituting inertialess new renewable sources of energy for 
inertiaful conventional generators critically needs to be accompanied by the deployment of 
synchronous condensers or synthetic inertia in Fiedler areas, while the need for inertia substitution 
is lower outside the Fiedler areas. Models similar to our dynamical model for the synchronous grid of continental Europe 
should prove to be helpful tools in planning for inertia deployment as the penetration of new renewables increases. 

\appendices \label{section:appendix}

 \begin{figure*}[h!]
\center
\includegraphics[width=\textwidth]{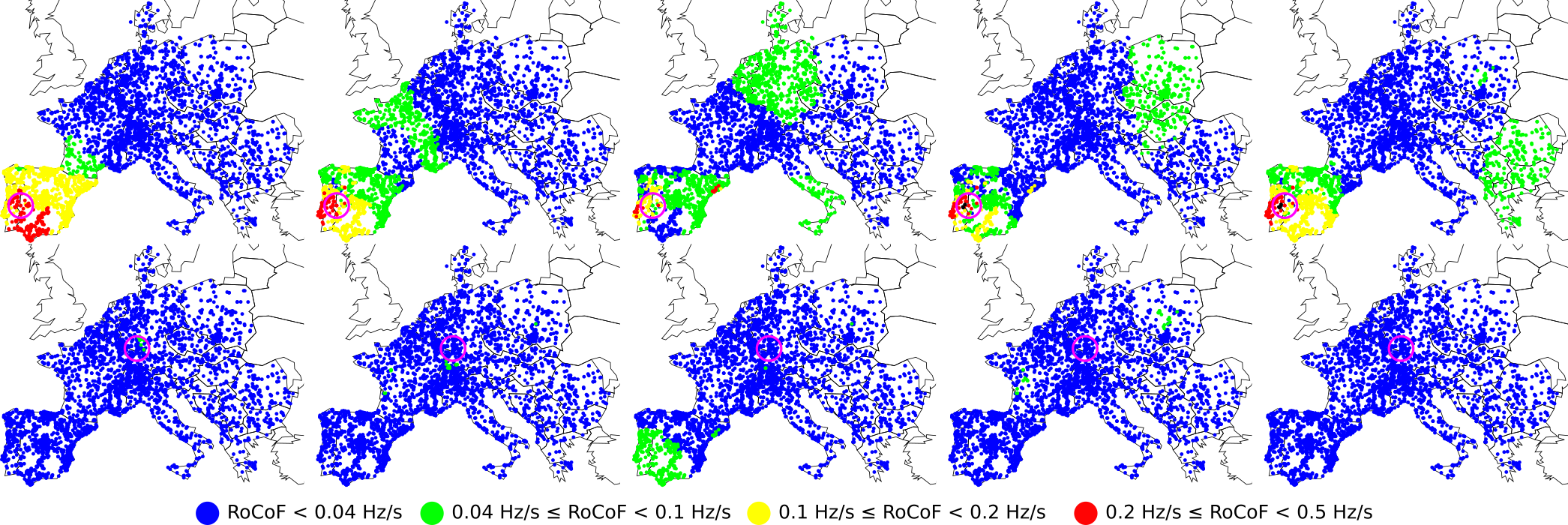}
\caption{Spatio-temporal evolution of local RoCoFs for two different power losses of $\Delta P=900$ MW.
in a moderate load (typical of a standard summer evening) configuration of the synchronous grid of continental Europe of 2018. 
The top four panels correspond to a fault in Spain and the bottom four to a fault in Western Germany. In both cases, 
the fault location is indicated by a purple circle. Panels correspond to 
snapshots over time intervals 0-0.5[s], 0.5-1[s], 1-1.5[s], 1.5-2[s] and 2-2.5[s] from left to right.}\label{fig:RoCoF_snapshots4}
\end{figure*}

\section{Disturbance Propagation for different faults or with different loads} \label{section:add_evolutions}

We present additional data complementing those in 
Figs.~\ref{fig:RoCoF_snapshots}, \ref{fig:RoCoF_snapshots2} and \ref{fig:RoCoF_snapshots3}
on disturbance propagation following a power loss. 

\begin{figure*}[h!]
\center
\includegraphics[width=\textwidth]{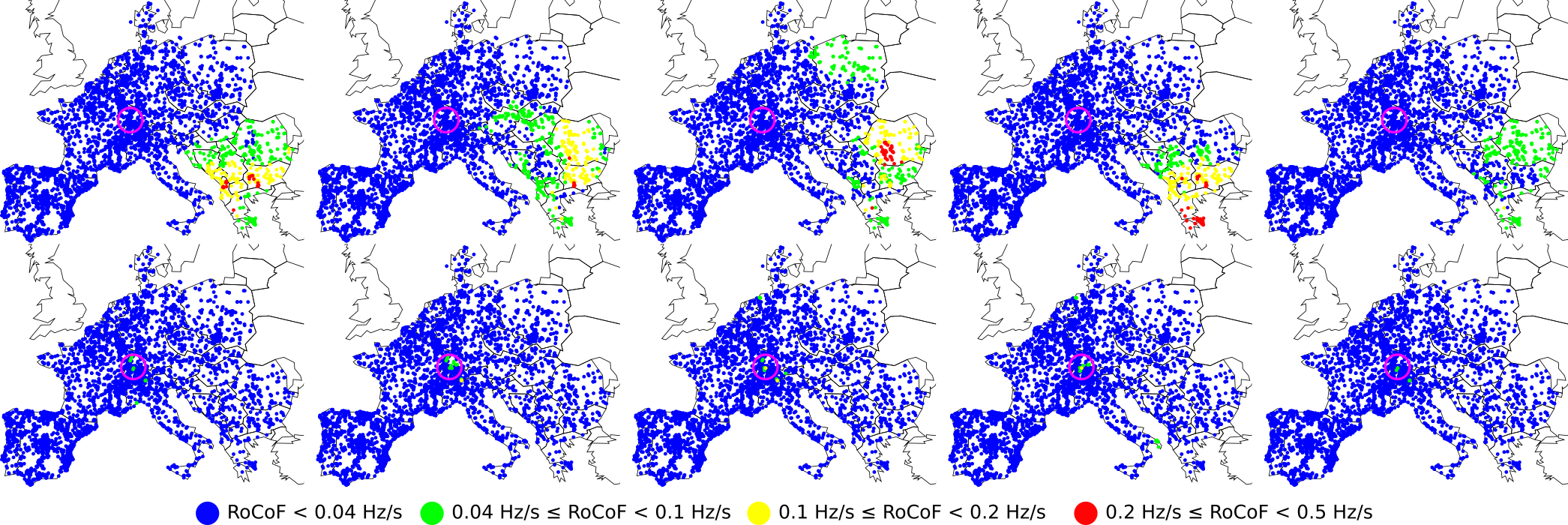}
\caption{Spatio-temporal evolution of local RoCoFs for the same two abrupt power losses of $\Delta P=900$  MW
as in Fig.~\ref{fig:RoCoF_snapshots} but with a higher load 
(typical of an early evening in winter) configuration. 
The top four panels correspond to a fault in Greece and the bottom four to a fault in Switzerland. In both cases, 
the fault location is indicated by a purple circle. Panels correspond to 
snapshots over time intervals 0-0.5[s], 0.5-1[s], 1-1.5[s], 1.5-2[s] and 2-2.5[s] from left to right.}\label{fig:RoCoF_snapshots5}
\end{figure*}

Fig.~\ref{fig:RoCoF_snapshots4} shows disturbance propagation following different power losses for the same 
load and the same European grid as in Fig.~\ref{fig:RoCoF_snapshots}. 
Fig.~\ref{fig:RoCoF_snapshots5} shows disturbance propagation following the same power losses 
as in Fig.~\ref{fig:RoCoF_snapshots}, but this time on a more loaded grid corresponding to a winter evening. 
Because of the larger load, more generators are working and therefore the total effective inertia on the grid is larger.
Disturbance propagation in both instances exhibit the same qualitative behavior as in  Fig.~\ref{fig:RoCoF_snapshots},
with quantitative differences arising at short times for Fig.~\ref{fig:RoCoF_snapshots5} with more inertia (as expected).

\section{A model of the synchronous grid of continental Europe}\label{section:model}

\subsection{Publicly available models}

There are only few publicly available models of part or all of the synchronous grid of continental Europe.
To the best of our knowledge, the first one was released by Zhou and Bialek~\cite{zhou2005approximate}
and later upgraded to incorporate the Balkans~\cite{hutcheon2013updated}. Other models include
ELMOD~\cite{leuthold2012large}, PEGASE~\cite{josz2016ac,zim11} and PyPSA-Eur~\cite{hor18}. These useful
power flow / optimal power flow models have not been extended to 
dynamical simulations and, except PyPSA-Eur, they lack bus geolocalization. 
We have constructed our own model to circumvent these two shortcomings. 
We briefly describe how this construction proceeded. 

\subsection{Geolocalized model for dynamical simulations of the continental European grid}\label{section:build}

We built our grid model from different publicly available databases under Creative Commons licenses. 
Similar building procedures were used in Refs.~\cite{zhou2005approximate,leuthold2012large}.
A similar model has recently been constructed, whose parameters do not seem to be publicly 
available~\cite{hewes2016development}.

\subsubsection{Geolocalization of buses and lines}

Wiegmans has extracted geolocalization data for the continental European grid from the 
ENTSO-E interactive map~\cite{wiegmans2016gridkit}. Our starting point is his database, which contains 
location and voltage of buses, identified as either generator or load buses, length and voltages for 
transmission lines and voltages for transformers. We determined the principal component
of that grid and discarded non-connected buses. The final network has 3809 buses connected by
4944 transmission lines.

\subsubsection{Electrical parameters of transmission lines}

Transmission lines operate at two different voltages of either 220 kV or 380 kV.
Within the lossless line approximation used in this manuscript, lines have purely imaginary admittances and are 
therefore characterized by their susceptance $B_{ij}$~\cite{bergen2000power}. They are given by  
\begin{align}
B_{ij}&=1\big/(X_{ij} l_{ij}) \, , 
\end{align}
where $l_{ij}$ is the length of the line (measured in kilometers) 
and $X_{ij}$ is its kilometric reactance. For the latter we use 
$X_{ij}=360 \, \rm m \Omega/$km for 220 kV lines and 
$X_{ij}=265 \, \rm m \Omega/$km for 380 kV lines. 
These values correspond to averages of those found in Ref.~\cite{siemens2014power}.

\subsubsection{Distribution of national loads}

Time series for national loads are available for  member countries of ENTSO-E~\cite{entsoe2015transparency}. 
For each country, we distribute those loads demographically over the set of consumer 
buses~\cite{zhou2005approximate,leuthold2012large}. 

Geographical population distributions are first determined from the GeoNames database~\cite{geonames}.
Second, the population of each town is distributed over all buses that are less than $d_{\max}=50$ km away
from it proportionally to their weight $w=1$ for 220 kV buses and $w=3$ for 380 kV buses.
This determines the effective population attributed to each bus. Third, 
the national load is distributed to each national bus in proportion to their attributed population. 
The validity of the procedure is at least partly confirmed by 
the strong correlation between population and load distributions in Italy reported in Ref.~\cite{zhou2005approximate}. 

\subsubsection{Conventional generators: capacity and dispatch}

\begin{figure}[h!]
\center
\includegraphics[width=0.45\textwidth]{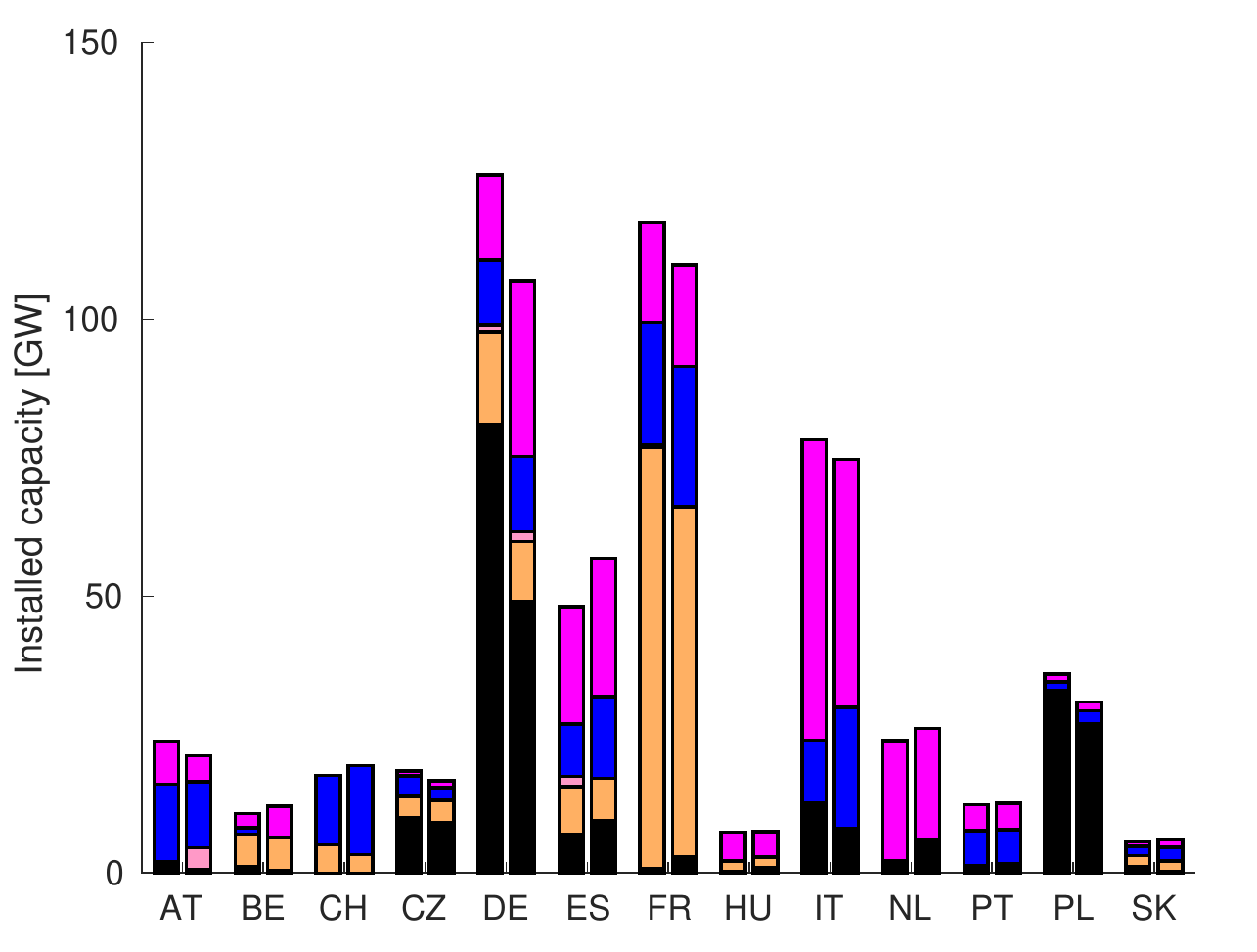}
\caption{Installed capacity according to our grid model (left column) and to Ref.~\cite{entsoe2015transparency} (right column)
for several European countries.}\label{fig:inst_cap}
\end{figure}

Wiegmans' database contains partial information on generator types and rated power~\cite{wiegmans2016gridkit}.
The missing generator data are obtained from the global energy observatory website~\cite{geopower}. 
Fig.~\ref{fig:inst_cap} compares the national installed capacities for some European countries in our model
to those listed in Ref.~\cite{entsoe2015transparency}.

Given distributed loads, we use an economic dispatch in the form of a DC optimal power flow 
to obtain power feed-in from  conventional generators.
Table~\ref{table:marginal_cost} gives the marginal costs $c_i$ that we use for different technologies.
\begin{table}[!h]
\caption{Marginal cost and inertia constant for different types of conventional generators.}\label{table:marginal_cost}
\center
\begin{tabular}{lcc}
\hline
 & $c_i$ [\$/MWh] & $H_i$ [s]\\
\hline
Hydro& 80 & 4\\
Nuclear& 16 & 6\\
Lignite& 16 & 6\\
Hard coal& 35 & 6\\
Gas& 100 & 6\\
Other& 7 & 3\\
\hline
\end{tabular}
\end{table}

\subsubsection{Conventional generators: inertia and damping coefficients}\label{section:M}

Rotating machines are characterized by their inertia constant $H_i$ which is 
the time over which the rated power of the generator provides a work equivalent to its kinetic 
energy~\cite{mac08,bergen2000power}. It is related to the inertia coefficient in \eqref{eq:swing1} via
\begin{equation}\label{eq:mihi}
m_i=\frac{2H_i P_i^{(0)}}{\omega_0}\, ,
\end{equation}
where $P_i^{(0)}$ is the rated power of the generator.
Table~\ref{table:marginal_cost} summarizes the values of inertia constants we use in our grid model~\cite{kundur1994power}.
Together 
with \eqref{eq:mihi} and the rated powers obtained in the previous paragraph, they determine the inertia
coefficients $m_i$ for each generator type. 

Damping coefficients $d_i$ are finally obtained for each generator type from Eq.~(5.24) and Table 4.3 in Ref.~\cite{mac08}.

\subsubsection{Frequency dependent loads}\label{section:freq_dep}

In the structure-preserving model we are using, loads are frequency-dependent~\cite{ber81}.
The load frequency/damping coefficient $d_i$ can be expressed as
\begin{equation}
d_i=\frac{\alpha P_{i}^0}{\omega_0}\,.
\end{equation}
with the load $P_{i}^0$ at nominal frequency $\omega_0$. 
The frequency-dependence of loads
has been experimentally investigated~\cite{welfonder1989investigations,o1996identification},
with reported values $\alpha\in 0.8-2$. In this manuscript, 
we use $\alpha = 1.5$. 

\section*{Acknowledgment}
\addcontentsline{toc}{section}{Acknowledgment}

This work has been supported by the Swiss National Science Foundation under an AP energy grant.


\begin{thebibliography}{10}
\providecommand{\url}[1]{#1}
\csname url@samestyle\endcsname
\providecommand{\newblock}{\relax}
\providecommand{\bibinfo}[2]{#2}
\providecommand{\BIBentrySTDinterwordspacing}{\spaceskip=0pt\relax}
\providecommand{\BIBentryALTinterwordstretchfactor}{4}
\providecommand{\BIBentryALTinterwordspacing}{\spaceskip=\fontdimen2\font plus
\BIBentryALTinterwordstretchfactor\fontdimen3\font minus
  \fontdimen4\font\relax}
\providecommand{\BIBforeignlanguage}[2]{{%
\expandafter\ifx\csname l@#1\endcsname\relax
\typeout{** WARNING: IEEEtran.bst: No hyphenation pattern has been}%
\typeout{** loaded for the language `#1'. Using the pattern for}%
\typeout{** the default language instead.}%
\else
\language=\csname l@#1\endcsname
\fi
#2}}
\providecommand{\BIBdecl}{\relax}
\BIBdecl

\bibitem{mac08}
J.~Machowski, J.~Bialek, and J.~R. Bumby, \emph{Power system dynamics:
  stability and control}, 2nd~ed.\hskip 1em plus 0.5em minus 0.4em\relax John
  Wiley \& Sons, 2008.

\bibitem{ulb14}
A.~Ulbig, T.~S. Borsche, and G.~Andersson, ``Impact of low rotational inertia
  on power system stability and operation,'' \emph{IFAC Proceedings Volumes},
  vol.~47, no.~3, pp. 7290--7297, 2014.

\bibitem{tie16}
P.~Tielens and D.~Van~Hertem, ``The relevance of inertia in power systems,''
  \emph{Renewable and Sustainable Energy Reviews}, vol.~55, pp. 999--1009,
  2016.

\bibitem{ulb15}
A.~Ulbig, T.~S. Borsche, and G.~Andersson, ``Analyzing rotational inertia, grid
  topology and their role for power system stability,''
  \emph{IFAC-PapersOnLine}, vol.~48, no.~30, pp. 541--547, 2015.

\bibitem{bev14}
H.~Bevrani, T.~Ise, and Y.~Miura, ``Virtual synchronous generators; a survey
  and new perspectives,'' \emph{Intl. Journal of Electrical Power and Energy
  Systems}, no.~54, pp. 244--254, 2014.

\bibitem{yan17}
J.~Yan, R.~Pates, and E.~Mallada, ``Performance tradeoffs of dynamically
  controlled grid-connected inverters in low inertia power systems,'' in
  \emph{IEEE 56th Annual Conference on Decision and Control}.\hskip 1em plus
  0.5em minus 0.4em\relax IEEE, 2017.

\bibitem{bor15}
T.~S. Borsche, T.~Liu, and D.~J. Hill, ``Effects of rotational inertia on power
  system damping and frequency transients,'' in \emph{IEEE 54th Annual
  Conference on Decision and Control}.\hskip 1em plus 0.5em minus 0.4em\relax
  IEEE, 2015, pp. 5940--5946.

\bibitem{poo17}
B.~K. Poolla, S.~Bolognani, and F.~D{\"o}rfler, ``Optimal placement of virtual
  inertia in power grids,'' \emph{IEEE Transactions on Automatic Control},
  vol.~62, no.~12, pp. 6209--6220, 2017.

\bibitem{pir17}
M.~Pirani, J.~W. Simpson-Porco, and B.~Fidan, ``System-theoretic performance
  metrics for low-inertia stability of power networks,'' in \emph{Decision and
  Control (CDC), 2017 IEEE 56th Annual Conference on}.\hskip 1em plus 0.5em
  minus 0.4em\relax IEEE, 2017, pp. 5106--5111.

\bibitem{bor18}
T.~S. Borsche and D{\"o}rfler, ``On placement of synthetic inertia with
  explicit time-domain constraints,'' \emph{arXiv:1705.03244}, 2017.

\bibitem{sia16}
M.~Siami and N.~Motee, ``Fundamental limits and tradeoffs on disturbance
  propagation in linear dynamical networks,'' \emph{IEEE Transactions on
  Automatic Control}, vol.~61, no.~12, pp. 4055--4062, 2016.

\bibitem{wol17}
J.~Wolter, B.~L\"unsmann, X.~Zhang, M.~Schr\"oder, and M.~Timme, ``Quantifying
  transient spreading dynamics on networks,'' \emph{arXiv:1710.09687}, 2017.

\bibitem{tam18}
S.~Tamrakar, M.~Conrath, and S.~Kettemann, ``Propagation of disturbances in ac
  electricity grids,'' \emph{Scientific Reports}, vol.~8, p. 6459, 2018.

\bibitem{bir17}
A.~B. Birchfield, T.~Xu, K.~M. Gegner, K.~S. Shetye, and T.~J. Overbye, ``Grid
  structural characteristics as validation criteria for synthetic networks,''
  \emph{IEEE Transactions on power systems}, vol.~32, no.~4, pp. 3258--3265,
  2017.

\bibitem{ber81}
A.~R. Bergen and D.~J. Hill, ``A structure preserving model for power system
  stability analysis,'' \emph{IEEE Transactions on Power Apparatus and
  Systems}, no.~1, pp. 25--35, 1981.

\bibitem{tyl18a}
M.~Tyloo, T.~Coletta, and P.~Jacquod, ``Robustness of synchrony in complex
  networks and generalized kirchhoff indices,'' \emph{Physical Review Letters},
  no. 120, p. 084101, 2018.

\bibitem{tyl18b}
M.~Tyloo, L.~Pagnier, and P.~Jacquod, ``The key player problem in complex
  oscillator networks and electric power grids: resistance centralities
  identify local vulnerabilities,'' \emph{to be published}, 2018.

\bibitem{jen00}
N.~Jenkins, R.~Allan, P.~Crossley, D.~Kirschen, and G.~Strbac, \emph{Embedded
  generation}.\hskip 1em plus 0.5em minus 0.4em\relax The Institution of
  Engineering and Technology, 2000.

\bibitem{entsoe2016frequency}
ENTSO-E, ``{Frequency Stability Evaluation Criteria for the Synchronous Zone of
  Continental Europe},''
  \url{https://docs.entsoe.eu/dataset/inertia-report-continental-europe/},
  2016.

\bibitem{zhou2005approximate}
Q.~Zhou and J.~W. Bialek, ``Approximate model of european interconnected system
  as a benchmark system to study effects of cross-border trades,'' \emph{IEEE
  Transactions on power systems}, vol.~20, no.~2, pp. 782--788, 2005.

\bibitem{hutcheon2013updated}
N.~Hutcheon and J.~W. Bialek, ``Updated and validated power flow model of the
  main continental european transmission network,'' in \emph{PowerTech, 2013
  IEEE Grenoble}.\hskip 1em plus 0.5em minus 0.4em\relax IEEE, 2013.

\bibitem{leuthold2012large}
F.~U. Leuthold, H.~Weigt, and C.~Von~Hirschhausen, ``A large-scale spatial
  optimization model of the european electricity market,'' \emph{Networks and
  spatial economics}, vol.~12, no.~1, pp. 75--107, 2012.

\bibitem{josz2016ac}
C.~Josz, S.~Fliscounakis, J.~Maeght, and P.~Panciatici, ``{AC power flow data
  in MATPOWER and QCQP format: iTesla, RTE snapshots, and PEGASE},''
  \emph{arXiv:1603.01533}, 2016.

\bibitem{zim11}
R.~Zimmerman, C.~Murillo-S\'anchez, and R.~T. Thomas, ``Matpower: Steady-state
  operations, planning and analysis tools for power systems research and
  education,'' \emph{IEEE Trans. on Power Systems}, vol.~26, pp. 12--19, 2011.

\bibitem{hor18}
J.~H\"orsch, F.~Hofmann, D.~Schlachberger, and T.~Brown, ``Pypsa-eur: An open
  optimisation model of the european transmission system,'' \emph{Energy
  Strategy Reviews}, vol.~22, pp. 207--215, 2018.

\bibitem{hewes2016development}
D.~Hewes, S.~Altschaeffl, I.~Boiarchuk, and R.~Witzmann, ``Development of a
  dynamic model of the european transmission system using publicly available
  data,'' in \emph{Energy Conference (ENERGYCON), 2016 IEEE
  International}.\hskip 1em plus 0.5em minus 0.4em\relax IEEE, 2016, pp. 1--6.

\bibitem{wiegmans2016gridkit}
B.~Wiegmans, ``{GridKit extract of ENTSO-E interactive map},''
  \url{https://doi.org/10.5281/zenodo.55853}, 2016.

\bibitem{bergen2000power}
A.~Bergen and V.~Vittal, \emph{Power Systems Analysis}, 2nd~ed.\hskip 1em plus
  0.5em minus 0.4em\relax Pearson/Prentice Hall, 2000.

\bibitem{siemens2014power}
``Power engineering guide,'' \url{http://siemens.com/energy/peg}, Siemens,
  2014.

\bibitem{entsoe2015transparency}
ENTSO-E, ``{ENTSO-E Transparency platform},''
  \url{https://transparency.entsoe.eu/}, 2015.

\bibitem{geonames}
Geonames, ``{Cites1000 data base},''
  \url{http://download.geonames.org/export/}.

\bibitem{geopower}
{Global energy observatory}, ``{GEO Power plants database},''
  \url{http://globalenergyobservatory.org/}.

\bibitem{kundur1994power}
P.~Kundur, N.~J. Balu, and M.~G. Lauby, \emph{Power system stability and
  control}.\hskip 1em plus 0.5em minus 0.4em\relax McGraw-hill New York, 1994.

\bibitem{welfonder1989investigations}
E.~Welfonder, H.~Weber, and B.~Hall, ``Investigations of the frequency and
  voltage dependence of load part systems using a digital self-acting measuring
  and identification system,'' \emph{IEEE Transactions on Power Systems},
  vol.~4, no.~1, pp. 19--25, 1989.

\bibitem{o1996identification}
J.~O'Sullivan and M.~O'Malley, ``Identification and validation of dynamic
  global load model parameters for use in power system frequency simulations,''
  \emph{IEEE Transactions on Power Systems}, vol.~11, no.~2, pp. 851--857,
  1996.

\end{thebibliography}
\end{document}